\documentclass[%
 reprint,
superscriptaddress,
 amsmath,amssymb,
 aps,
 onecolumn,
 longbibliography,
]{revtex4-2}
\usepackage{tikz}

\usepackage{graphicx}
\usepackage{dcolumn}
\usepackage{bm}
\usepackage{tabularx}
\usepackage{caption}
\usepackage[labelformat=simple]{subcaption}\usepackage[section]{placeins}
\usepackage{flafter}
\usepackage{wasysym}
\usepackage{xcolor}

\begin{document}


\title{Segregation models for density-bidisperse granular flows}

\author{Yifei Duan}
\affiliation{
Department of Chemical and Biological Engineering, Northwestern University, Evanston, Illinois 60208, USA, \looseness=-1
}%
\author{Paul B. Umbanhowar}
\affiliation{
Department of Mechanical Engineering, Northwestern University, Evanston, Illinois 60208, USA
}%
\author{Julio M. Ottino}
\affiliation{
Department of Chemical and Biological Engineering, Northwestern University, Evanston, Illinois 60208, USA, \looseness=-1
}%
\affiliation{
Department of Mechanical Engineering, Northwestern University, Evanston, Illinois 60208, USA
}%
\affiliation{
Northwestern Institute on Complex Systems (NICO), Northwestern University, Evanston, Illinois 60208, USA, \looseness=-1
}%
\author{Richard M. Lueptow}
 \email{Corresponding author: r-lueptow@northwestern.edu}
 \affiliation{
Department of Chemical and Biological Engineering, Northwestern University, Evanston, Illinois 60208, USA, \looseness=-1
}%
\affiliation{
Department of Mechanical Engineering, Northwestern University, Evanston, Illinois 60208, USA
}%
\affiliation{
Northwestern Institute on Complex Systems (NICO), Northwestern University, Evanston, Illinois 60208, USA, \looseness=-1
}%

\date{\today}

\begin{abstract}
Individual constituent balance equations are often used to derive expressions for species specific segregation velocities in flows of dense granular mixtures.
We propose a semi-empirical expression for the interspecies momentum exchange in density bidisperse granular flows as an extension of ideas from kinetic theory and compare it to a previous viscous drag approach that is analogous to particles settling in a fluid. The proposed model expands the range of the granular kinetic theory from short-duration binary collisions to the multiple enduring contacts characteristic of dense shear flows and incorporates the effects of particle friction, concentration ratio, and local flow conditions.
The segregation velocities derived from the momentum balance equation using both interspecies drag models match the downward and upward segregation velocities of heavy and light particles obtained from DEM simulations through the flowing layer depth for different density ratios and constituent concentrations in confined shear flows. Predictions of the kinetic theory inspired approach are additionally compared to results from free surface heap flow simulations, and, again, a close match is observed.
\end{abstract}

\maketitle


\section{Introduction}
Dense granular materials, when sheared, tend to segregate by particle size, density, shape or friction coefficient, which can be problematic in many industries due to its impact on product quality and uniformity \cite{gray2005theory,meier2007dynamical,fan2011theory,tunuguntla2014mixture,xiao2017controlling,zhao2018simulation,xiao2019continuum,umbanhowar2019modeling}.
As a result, the segregation behavior of sheared granular mixtures is of interest in a wide variety of fields from both fundamental and applied standpoints.

We focus on bidisperse granular mixtures of particles having the same size but different densities such that segregation is driven by a  ``buoyant force" mechanism \cite{khakhar1997radial,khakhar1999mixing}. 
ln this paper, we propose a model based on the Kinetic Thoery of Granular Flow (KTGF) for the interspecies drag between heavy and light particles in density bidisperse granular flows and use the model to derive an expression for the segregation velocity of each species by considering the balance between the net buoyant force and the interspecies drag force.
We then compare this new drag model to a previous viscous drag model \cite{tripathi2011numerical,tripathi2013density} that is analogous to the viscous drag force acting on a particle settling in a fluid.

To model the segregation process for both approaches, we assume that two different particle species can be treated as interpenetrable continua such that each constituent has its own partial pressure, consistent with previous studies \cite{gray2005theory,fan2011theory,tunuguntla2014mixture}. The imbalance between a species' partial pressure and its body force results in segregation of the two species.
%
A linear relation between the drag force on a particle and its segregation velocity has been used in several previous gravity-driven segregation models for size or density segregation \cite{gray2005theory,tripathi2013density,tripathi2011numerical,gajjar2014asymmetric,gray2015particle,tunuguntla2014mixture,marks2012grainsize} based on either the similarity of kinetic sieving in granular segregation with fluid percolation through a porous material or an analogy to the drag force on a sphere settling in a fluid. 
The linear drag model has been combined with the constituent mass and momentum balances to derive a general multicomponent theory for segregation \cite{gray2018particle}. 
However, \textcite{weinhart2013discrete} demonstrated that the coefficient fitted from the linear drag model varies with time, indicating that a simple linear drag law fails to describe the segregation behavior.
Recent DEM simulations also suggest that other factors, such as the local pressure \cite{fry2018effect,liu2017transport} and the species concentration ratio \cite{van2015underlying,jones2018asymmetric}, affect the segregation velocity, which again raises questions concerning the validity of a linear drag model.

As an alternative to the linear drag model, it is natural to consider the KTGF to model the momentum transfer between two segregating species. The KTGF provides a drag model connecting stresses and velocities in a granular mixture while accounting for complex particle interactions \cite{gidaspow1986hydrodynamics, syamlal1987particle,huilin2003hydrodynamics,huilin2007investigation,iddir2005analysis,iddir2005modeling,chao2012investigation}, 
thereby yielding a drag model that takes particle density and diameter into consideration at the particle level. However, there are two issues with using the KTGF approach to model dense flows. First, the stress generation mechanism in the KTGF is based on short-duration, binary collisions typical of dilute particle flows as opposed to multiple, enduring contacts typical of the dense granular flows considered here. 
Second, the KTGF approach does not include physics known to be critical to segregation in dense granular flows such as the dependence of segregation on the local pressure \cite{fry2018effect}. 
Nevertheless, a potential path toward a robust drag model for dense flows lies in modifying the KTGF drag model based on empirical results for density segregation in dense granular flows.

In this paper, we derive expressions for the segregation velocity in density bidisperse granular flow by combining the particle interspecies drag and the equilibrium momentum balance equation from mixture theory \cite{gray2005theory} using two different approaches to model the interspecies drag.
One approach integrates particle-particle collisional force within the framework of KTGF, while the other adopts the form of a particle-fluid viscous force based on a modified Stokes law.
Both models are described in Section \ref{section2}.
The effects of the local inertial number, particle friction, and relative species concentration proposed in the KTGF drag model are tested separately by DEM simulations in Section \ref{section3}.
In Section \ref{section4}, the viscous drag coefficient is determined and the granular flow rheology for calculating the pseudo-viscosity of the mixture is discussed.  
In Section \ref{section5}, results of both segregation models are compared with DEM simulations for flows under more general conditions. Conclusions are given in Section \ref{section6}.

\section{drag models and segregation velocity}
\label{section2}

%
For dense granular flows, the total solids volume fraction, $\phi_{solid}=\sum \phi_i$, where $\phi_{i}$ represents the local volume fraction of each solid species (here $i=h$ for heavy particles and $i=l$ for light particles), is around 0.6. The local constituent concentration is $c_i=\phi_i/\phi_{solid}$ and the solid density is $\rho_{solid}=\sum \rho_i c_i$, where $\rho_i$ is the particle material density.
The gradient of the lithostatic pressure $P$ is 
\begin{equation}
	\frac{\partial P}{\partial z}=-\rho_{solid} \phi_{solid} g,
	\label{gradient}
\end{equation}
where $z$ is the vertical coordinate (assuming negative $z$ is the gravity direction) and $g$ is the acceleration due to gravity.

The partial pressure $P_i$ is determined by partitioning the lithostatic pressure $P$ induced by gravity among the two constituents such that $P=\sum P_i$. 
For particles that only differ in density, we assume that the proportion of the hydrostatic load carried by each species equals its concentration $c_i$ regardless of the density difference as proposed by \textcite{marks2012grainsize} and \textcite{tunuguntla2014mixture}. 
The momentum balance for each species can be written as \cite{tunuguntla2014mixture}
\begin{equation}
    \frac{\partial}{\partial t} (\rho_i \phi_i \pmb u_i)  + \pmb\nabla \cdot (\rho_i \phi_i \pmb u_i \otimes \pmb u_i)  =-c_i  \pmb\nabla{P}+\rho_i \phi_i \pmb g +\pmb \beta_i,
    \label{momentum}
\end{equation}
where $\otimes$ is the dyadic product, $t$ is time, $\pmb u_i$ is the vector velocity, and $\pmb \beta_i$ is the interspecies drag vector. 
(Eq.~(\ref{momentum}) is similar to Eq.~(2.5) in \cite{tunuguntla2014mixture}, noting that the intrinsic density $\rho^{i *}$ in \cite{tunuguntla2014mixture} is the bulk density of particles in a non-mixed state, which is equal to $\rho_i \phi_{solid}$, and that $\rho^i=c^i \rho^{i *}$  is equivalent to $\rho_i \phi_i$ used here.)
Assuming that the vertical acceleration terms are negligible \cite{gray2006particle,tunuguntla2014mixture}, which is reasonable for the relatively slow segregation in a typical granular flowing layer, the momentum conservation equation in the $z$-direction can be simplified using Eq.~(\ref{gradient}) to 
\begin{equation}
   (\rho_{solid}-\rho_i) \phi_i g + \beta_i=0.
   \label{dragmomentum}
\end{equation}
Equation~(\ref{dragmomentum}) is analogous to a simple force balance between the net buoyant force $(\rho_{solid}-\rho_i)\phi_i g$ and the interspecies drag force $\beta_i$.

\subsection{KTGF based model}

To determine constituent velocities from Eq.~(\ref{dragmomentum}), $\beta_i$ must be expressed as a function of the velocity difference between the two species, and the KTGF offers a means to do this. Changes to the KTGF are needed, however, because it was developed to model rapid flow of dilute granular materials  based on instantaneous binary collisions \cite{jenkins1983theory,lun1984kinetic}.
This is not the case in the dense flow regime where enduring contacts dominate and many adjacent particles are part of contact force chains \cite{liu1995force}.
Correlations in motion and force due to the dense contact network reduce the collisional energy dissipation \cite{jenkins2006dense}. 
Substantial effort has been expended to extend the applicability of the KTGF to the dense regime, mainly by modifying the expression for the collisional energy dissipation rate to account for the effects of sustained contacts \cite{jenkins2007dense, jenkins2010dense, vescovi2014plane, duan2017modified,duan2019new}.
In fact, an augmented KTGF approach has been proposed for segregation problems, but it is limited to small size and density differences \cite{larcher2013segregation,larcher2015evolution}. 
In addition, several stand-alone drag models based on the KTGF have been proposed for dilute granular flows \cite{gidaspow1986hydrodynamics, syamlal1987particle,huilin2003hydrodynamics,huilin2007investigation,iddir2005analysis,iddir2005modeling,chao2012investigation}. 

Here we follow an approach similar to one of these KTGF drag models \cite{chao2012investigation} by assuming that the long-duration collisional drag in dense flows follows similar physics to the short-duration collisional drag in dilute flows, but with a much longer contact time. This can be represented by correction coefficients modifying the well-known solid-solid drag model proposed by \textcite{syamlal1987particle} that has been successfully used for fluidized bed simulations \cite{syamlal1993mfix,owoyemi2007cfd,hernandez2015fully}. 
We choose the Syamlal drag model \cite{syamlal1987particle} over other models for two reasons. First, within the KTGF framework the interspecies drag at the continuum level is derived by integrating the momentum transfer between two particles over the velocity space.
Thus, the dependence of drag on particle density has a physical basis at the particle level. Second, the drag model is derived by assuming all the particles of each type have identical velocities such that velocity fluctuations are not considered, making a closed-form expression for the drag easy to obtain.

The Syamlal expression for the local drag $\pmb \beta_i$ between two constituents $i$ and $j$ is 
\begin{equation}
   \pmb \beta_i=-\bigg[ \frac{ 3 (1+e) }{ 2\pi (\rho_i d_i^3+\rho_j d_j^3)}\bigg( \frac{\pi}{2} + \frac{\mu \pi^2}{8}  \bigg) \bigg] g_{ij}  \rho_i \rho_j \phi_i \phi_j   (d_i+d_j)^2  |\pmb u_i -\pmb u_j | (\pmb u_i -\pmb u_j),
    \label{eq771}
\end{equation}
where $g_{ij}$ is the radial distribution function for a collision between particle $i$ and $j$ \cite{lebowitz1964exact}, $\mu$ is the interparticle friction coefficient, and $e$ is the restitution coefficient. For density segregation, $d_i=d_j$, and $g_{ij}$ becomes a constant if we assume constant total solid volume fraction $\phi_{solid}$.
The bulk velocity in the vertical direction of the combined constituents is $w_{bulk}=c_i w_i +c_j w_j$ based on volume conservation.
The segregation, or percolation velocity $w_{p,i}$, for species $i$ is defined as the velocity difference between the constituent velocity and the bulk velocity in the $z$ direction:
\begin{equation}
	w_{p,i}=w_i-w_{bulk}=c_j( w_i - w_j).
	\label{bulkv}
\end{equation}
Using Eq.~(\ref{bulkv}) in Eq.~(\ref{eq771}) for bidisperse granular mixtures of heavy and light particles having the same diameter $d$, the interspecies drag in the segregation ($z$) direction is
\begin{equation}
    \beta_{i}=-\bigg[ { 3 (1+e) }\bigg( 1 + \frac{\mu \pi}{4}   \bigg) g_{ij} \bigg]  \bigg[  \frac{\rho_h \rho_l \phi_i }{(\rho_h+\rho_l)\phi_j} \phi_{solid}^2 \frac{|w_{p,i}|}{d}w_{p,i} \bigg].
    \label{eq77}
\end{equation}

The Syamlal model in Eq.~(\ref{eq77}) is derived by assuming that the particles of each type have identical velocities, so velocity fluctuations (i.e., the granular temperature) are not explicitly considered in the expression. We choose the Syamlal model over granular temperature dependent drag models knowing that in dense granular flows the Maxwellian velocity distribution is no longer valid and that the pressure-shear state, rather than the granular temperature, determines flow behavior \cite{pouliquen2006flow}. 
However, Eq.~(\ref{eq77}) indicates that the two particle species would segregate even without the presence of shear, which, of course, does not occur.
There are several ways to rationalize this. For example, \textcite{gera2004hydrodynamics} extended the Syamlal model to dense fluidized beds by adding a ``hindrance effect'' term related to friction and pressure, such that the momentum exchange has to exceed a threshold to drive the segregation.
Here we take a heuristic approach based on the $\mu(I)$ rheology \cite{midi2004dense,da2005rheophysics,jop2006constitutive}, accounting for local flow conditions by introducing the inertial number $I$, defined as 
\begin{equation}
	I=\dot \gamma d \sqrt{ \frac{\rho_{solid}}{P}},
	\label{inert}
\end{equation}
where $\dot \gamma$ is the local shear rate.
A small value of $I$ (small $\dot \gamma$ and/or large $P$) corresponds to the quasistatic regime where granular flows behave more like deforming solids. 
Conversely, a large value of $I$ (large $\dot \gamma$ and/or small $P$) corresponds to the collisional regime where granular flows behave more like fluids.
Based on simulation data that we describe in detail later in this paper, we propose a semi-empirical interspecies drag model of the form
\begin{widetext}
\begin{equation}
    \beta_i=\underbrace{-B(\mu) }_{\substack{\text{enduring contacts} } }  \overbrace{ \bigg[ \frac{\rho_h \rho_l \phi_i }{(\rho_l+\rho_h) \phi_j }  \phi_{solid}^2  \frac{|{w_{p,i}|}}{d } w_{p,i} \bigg]}^{\text{KTGF}} \underbrace{\frac{1}{I^2}}_{\substack{\text{local flow} }}  \overbrace{\sqrt{\frac{\phi_h}{\phi_l}}}^{\substack{\text{local } \\ \text{concentration ratio}}},
    \label{eq8}
\end{equation}
\end{widetext}
for density-bidisperse granular flows with inertial number $I<0.5$, corresponding to dense granular flow, and particle concentration, $0.1\leq c_i \leq0.9$.

Among the four multiplicative terms in Eq.~(\ref{eq8}), the KTGF term is inherited from the Syamlal model in Eq.~(\ref{eq77}) and reflects the momentum transfer between two particles with different masses, taking into consideration density and diameter at the particle level, while the other three terms are semi-empirical modifications.  
$B(\mu)$ replaces the first term in square brackets in Eq.~(\ref{eq77}) to account for enduring contacts in dense flows.
In dense granular mixtures short-duration collisions are unlikely to play as important of a role as in the dilute flows upon which Eq.~(\ref{eq77}) for KTGF is based. As a result, we assume the empirical function $B(\mu)$ is independent of $e$.
Previous studies show that the segregation velocity $w_{p,i}$ is proportional to the inertial number $I$ at constant $\beta_i$ \cite{fry2018effect,liu2017transport}, so the inertial number appears as $1/I^2$ in Eq.~(\ref{eq8}) to account for the local flow conditions.
Finally, the last term $\sqrt{\phi_h/\phi_l}$ accounts for the nonlinear dependence of segregation on particle concentration (i.e., heavy particles among many light particles segregate faster than light particles among many heavy particles \cite{jones2018asymmetric}), as it increases the drag with increasing ${{\phi_h}/{\phi_l}}$. 
Detailed justifications for these three modifying terms are provided in Section \ref{section3}.

The ultimate goal of this paper is to develop an accurate model for the segregation velocities of the light and heavy particle species in density bidisperse granular flows. Such an expression can be readily derived by substituting the expression for the interspecies drag $\beta_i$ from Eq.~(\ref{eq8}) into the equilibrium momentum balance of Eq.~(\ref{dragmomentum}). For light particles,
\begin{equation}
    \underbrace{(\rho_{solid}-\rho_l )g\phi_l}_{\text{net buoyant forces}} -  \overbrace{B(\mu)\frac{\rho_h \rho_l \phi_l }{(\rho_l+\rho_h) \phi_h } \frac{\phi_{solid}^2}{d} (\frac{w_{p,l}}{ I })^2\sqrt{\frac{\phi_h}{\phi_l}} }^{\text{interspecies drag $\beta_l$, Eq.~(\ref{eq8})}}=0,
    \label{eqseg}
\end{equation}
such that $w_{p,l}$ takes the form
\begin{equation}
    w_{p,l}=\bigg[ \frac{gd}{B(\mu) \phi_{solid}}   \big( R_\rho-\frac{1}{R_\rho})  \sqrt{\frac{c_l}{c_h}}     \bigg]^{1/2}  (1-c_l) I,
    \label{final}
\end{equation}
noting that $\phi_h/\phi_l=c_h/c_l$ and that $R_\rho=\rho_h/\rho_l$ is the density ratio. For heavy particles, a similar approach yields
\begin{equation}
    w_{p,h}=-\bigg[ \frac{gd}{B(\mu) \phi_{solid}}    \big( R_\rho-\frac{1}{R_\rho})   \sqrt{\frac{c_l}{c_h}}  \bigg]^{1/2}  (1-c_h) I.
    \label{finalh}
\end{equation}
Note that Eqs.~(\ref{final}) and (\ref{finalh}) apply to density-bidisperse granular flows with $I<0.5$ (dense flows), and particle concentrations, $0.1\leq c_i \leq0.9$ (as opposed to situations approaching a single intruder particle at concentrations outside this range).
Equations~(\ref{final}) and (\ref{finalh}) indicate that $w_{p,i} \propto \sqrt{R_\rho-\frac{1}{R_\rho}}$, which is similar to the relation $w_{p,i} \propto (\sqrt{R_\rho}-\frac{1}{\sqrt{R_\rho}})$ observed in monodisperse systems with several heavy intruder particles \cite{liu2017transport}. 
Equations~(\ref{final}) and (\ref{finalh}) also resemble the general form of the segregation velocity given by \textcite{fry2018effect} 
\begin{equation}
	w_{p,i}=\sqrt{gd}f(R_\rho)(1-c_i)I,
\end{equation}
by making 
\begin{equation}
	f(R_\rho)=\bigg[ \frac{1}{B(\mu) \phi_{solid}}    \big( R_\rho-\frac{1}{R_{\rho}} \big)   \sqrt{\frac{c_l}{c_h} }    \bigg]^{1/2}.
\end{equation}

These similarities suggest that the drag model proposed in Eq.~(\ref{eq8}) is reasonable and consistent with previous research.

\subsection{Modified stokes law based model}

For purposes of comparison, we also consider an alternative approach based on a viscous drag model \cite{tripathi2011numerical,tripathi2013density}. As an analogy to the viscous force acting on a particle settling in a fluid, \textcite{tripathi2013density} propose a modified Stokes viscous drag force, $F_{visc}=\epsilon \pi \eta d w_{p,i}$, where $\epsilon$ is an empirically determined drag coefficient that depends on the solid volume fraction $\phi_{solid}$, and the pseudo-viscosity $\eta$ of the granular mixture is assumed to be the origin of the drag. 
This differs from the KTGF approach where the drag is assumed to be a direct consequence of particle-particle interactions. The conceptual difference leads to different expressions for the segregation velocity. 
Using the expression for the modified Stokes drag force, the interspecies drag can be expressed as
\begin{equation}
	\beta_{visc,i}=\frac{F_{visc}}{\frac{1}{6}\pi d^3/\phi_i}=6\epsilon\eta \phi_{i} w_{p,i}/d^2.
	\label{stokesdrag}
\end{equation}
Substituting this expression into the momentum balance equation, Eq. (\ref{dragmomentum}), the segregation velocities take the form   
\begin{equation}
	w_{visc,l}=\frac{gd^2}{6\epsilon\eta}(\rho_h-\rho_l)(1-c_l) .
	\label{stokesvell}
\end{equation}
and 
\begin{equation}
	w_{visc,h}=-\frac{gd^2}{6\epsilon\eta}(\rho_h-\rho_l)(1-c_h) .
	\label{stokesvelh}
\end{equation}
Comparison of these two equations with Eqs.~(\ref{final}) and (\ref{finalh}) for the modified KTGF drag model are enlightening. Both forms depend on gravity, and particle size, density, and concentration. While the KTGF form is slightly more complicated, it only requires one empirical function, $B(\mu)$. On the other hand, the challenge in applying the viscous segregation model is to empirically determine the drag coefficient $\epsilon(\phi_{solid})$ and the pseudo-viscosity $\eta$ of the granular mixtures.
Both forms depend on the inertial number $I$, either directly for the KTGF model or indirectly through the determination of $\eta$ for the viscous drag model, as will be shown in Section \ref{section4}.

\section{Confirming the modified KTGF drag model}
\label{section3}

\begin{figure}
    \captionsetup{justification=raggedright,singlelinecheck=false}
    \centerline{\includegraphics[scale=0.5]{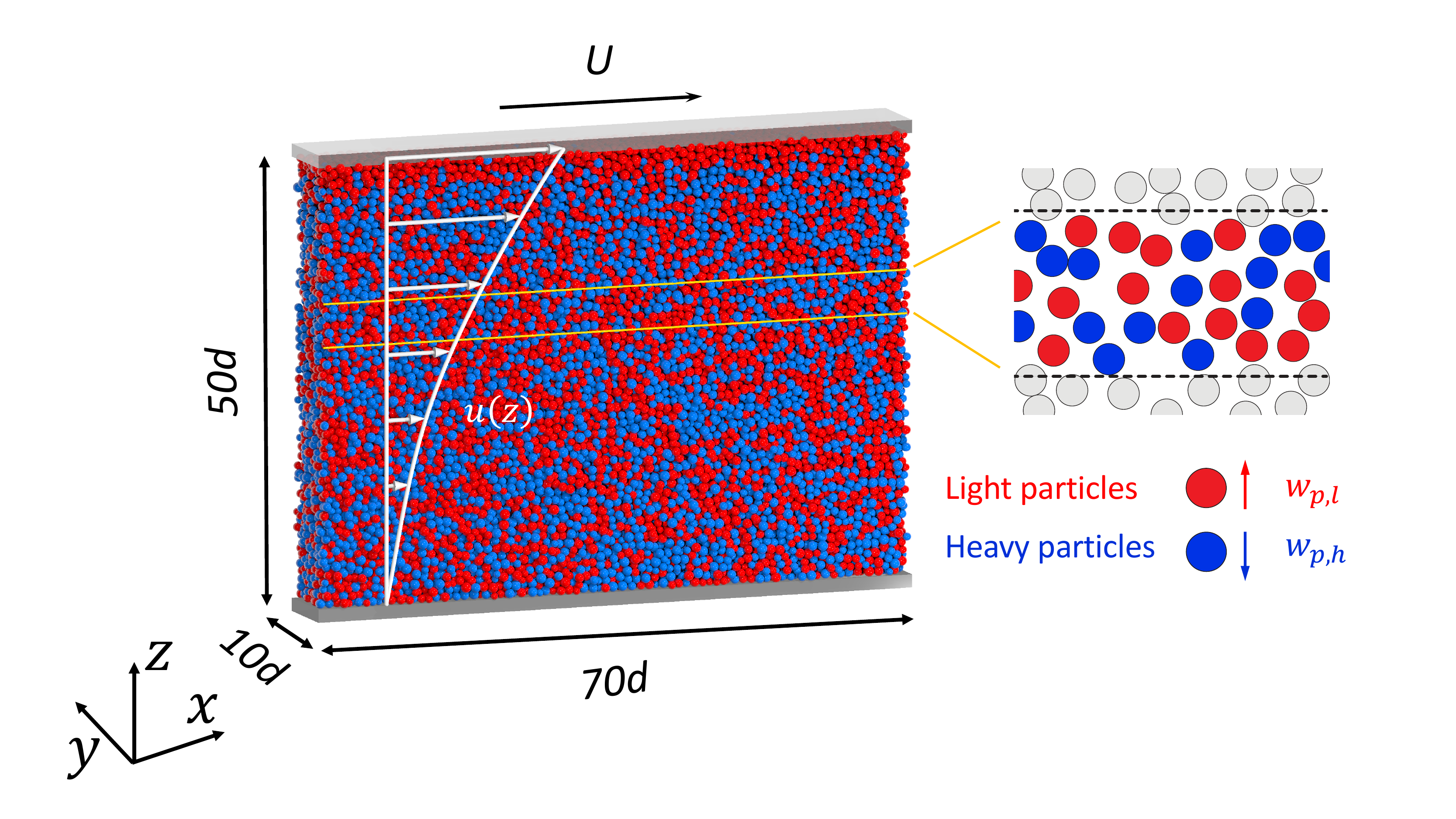}}
    \caption{Schematic of DEM simulation setup.}
    \label{scheme}
\end{figure}

In this section, we validate the interspecies drag model in Eq.~(\ref{eq8}) by performing DEM simulations under a variety of conditions. 
Direct measurement of $\beta_i$ from DEM simulations is difficult since both the segregation driving forces and the drag forces result from interparticle collisional forces; it is hard to distinguish them at the particle level. 
A better way to determine $\beta_i$ is to utilize momentum conservation at force equilibrium, Eq.~(\ref{dragmomentum}), in which ${\beta_i}$ equals $-(\rho_{solid}-\rho_i)\phi_i g$, a quantity that is easily measured locally in a granular flow. 
This is equivalent to validating the segregation model in Eqs.~(\ref{final}) and (\ref{finalh}), noting that all of the variables in these equations except $B(\mu)$ are known (i.e., $R_\rho$, $\phi_{solid}$, $d$) or can be calculated from the simulation (i.e., $w_{p,i}$, $I$). Hence, the problem comprises finding $B(\mu)$ under a variety of local flow conditions, $I$, and local concentration ratios, $c_h/c_l$.
\begin{figure}
\captionsetup{justification=raggedright}
    \centerline{\includegraphics[scale=0.5]{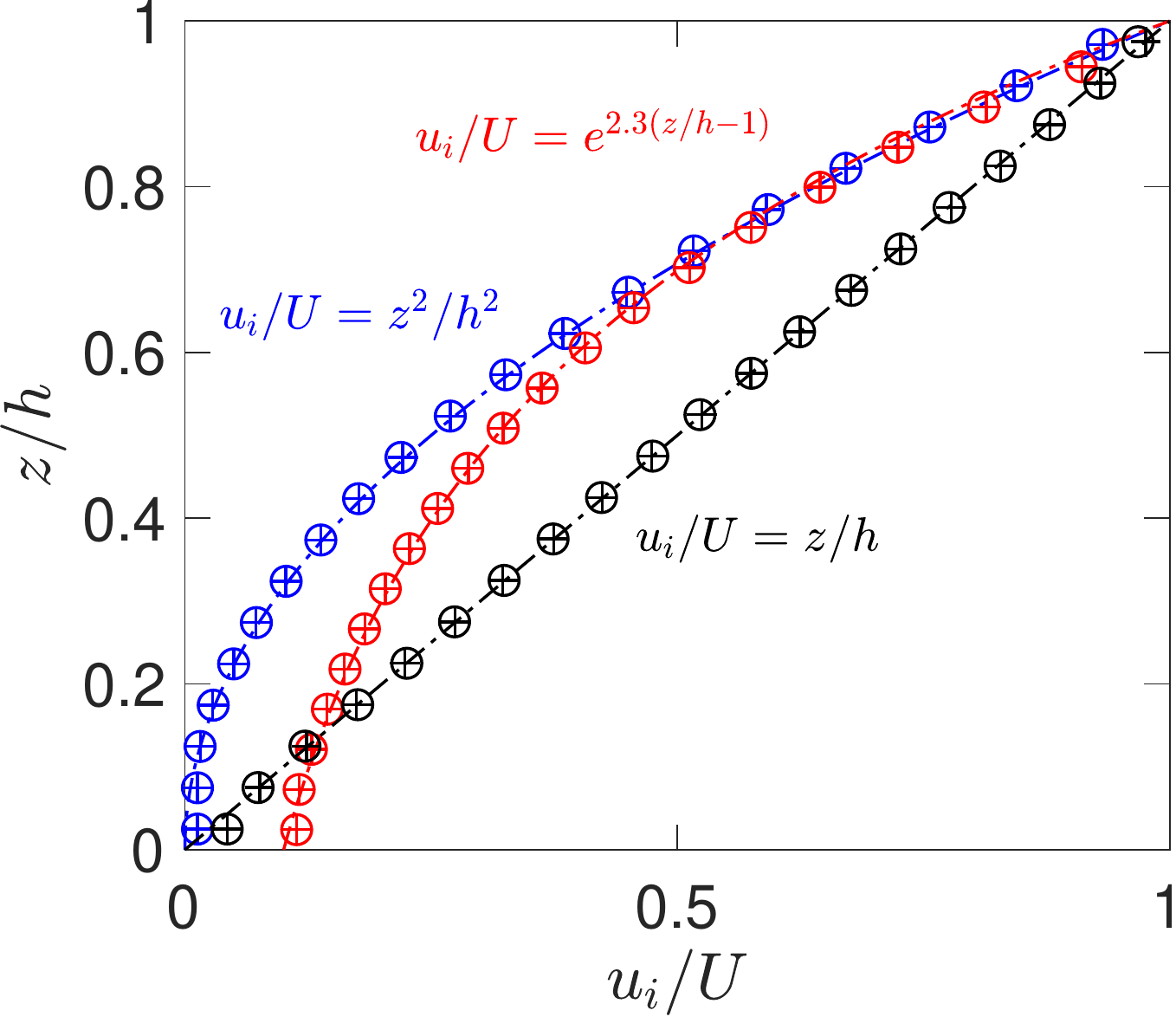}}
    \caption{Instantaneous streamwise velocity $u_i$ of each constituent for mixed light ($\textcolor{black}{+}$)  and heavy ($\textcolor{black}{\Circle}$) particles averaged in each horizontal layer $0.1~$s after shear onset.  
Dashed curves represent the velocity profiles imposed using the forcing specified in Eq.~(\ref{stable}).    
($d_h=d_l=4~$mm, $c_h=c_l=0.5$, $\mu=0.2$, $e=0.9$, and $R_\rho=4$.) }
    \label{uprofile}
\end{figure}

\begin{table}[h]
   \caption{Simulation conditions.}
   \label{table1}	
   \def\arraystretch{1.4}
   \begin{tabular}{p{3cm} p{6cm} }
   \toprule
   $ d$ & $4~$mm\\
   $ R_\rho=\rho_h/\rho_l$ & $1 -10$ \\
      $\mu$ & $0-0.6$ \\
   $c_h$, $c_l$ & $0.1- 0.9$ \\
    $e$ & $0.2- 0.9$ \\
        $U$ & $1- 5 ~\text{m/s}$ \\
    $u$ & $U{z}/{h}$, $Uz^2/h^{2}$, ${U} e^{2.3(z/h-1)}$\\
    $\dot\gamma$ & $U/h$, $2Uz/h^{2}$, $2.3Ue^{2.3(z/h-1)}/h$\\
    
    \botrule
   \end{tabular}
          \label{table1}
          
\end{table}

\begin{figure}
   \captionsetup[subfigure]{labelformat=empty}
    \captionsetup{justification=raggedright}
    \begin{subfigure}{0.45\columnwidth}
    \includegraphics[scale=0.494]{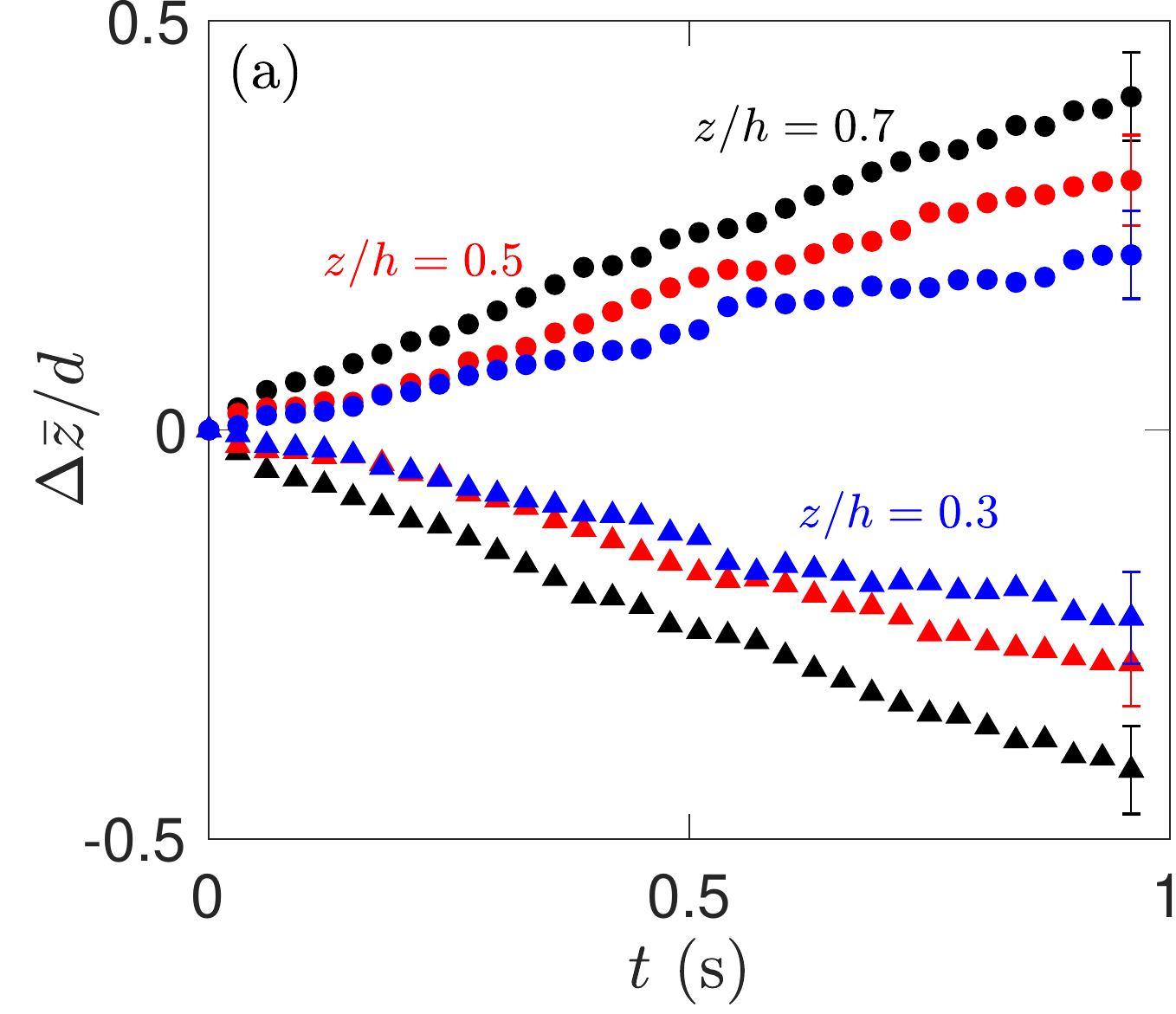}
    \subcaption{}\label{linear}
     \end{subfigure}
          \begin{subfigure}{0.45\columnwidth}
    \includegraphics[scale=0.494]{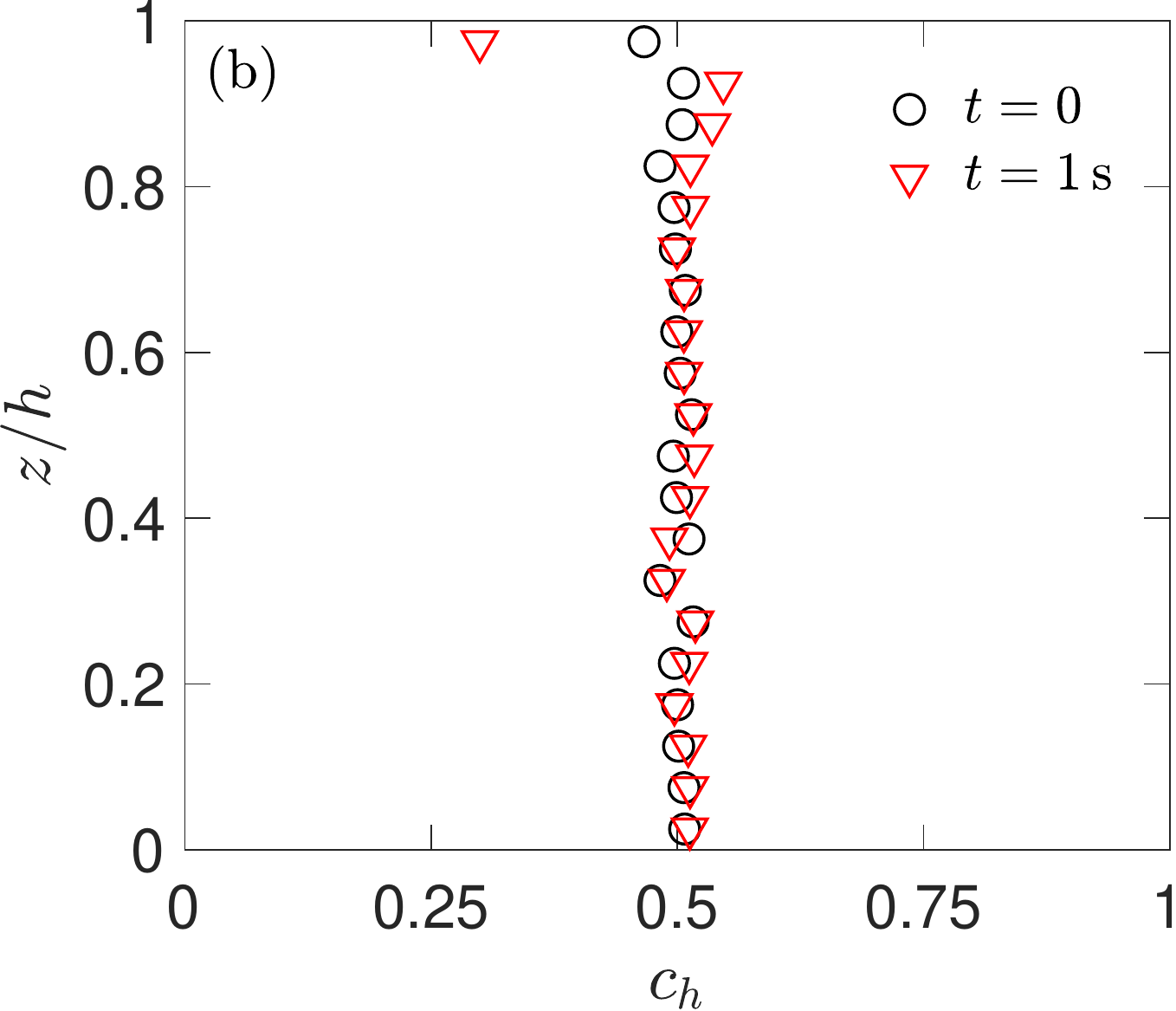}
    \subcaption{}\label{concentration}
    \end{subfigure}
            \begin{subfigure}{0.45\columnwidth}
    \includegraphics[scale=0.494]{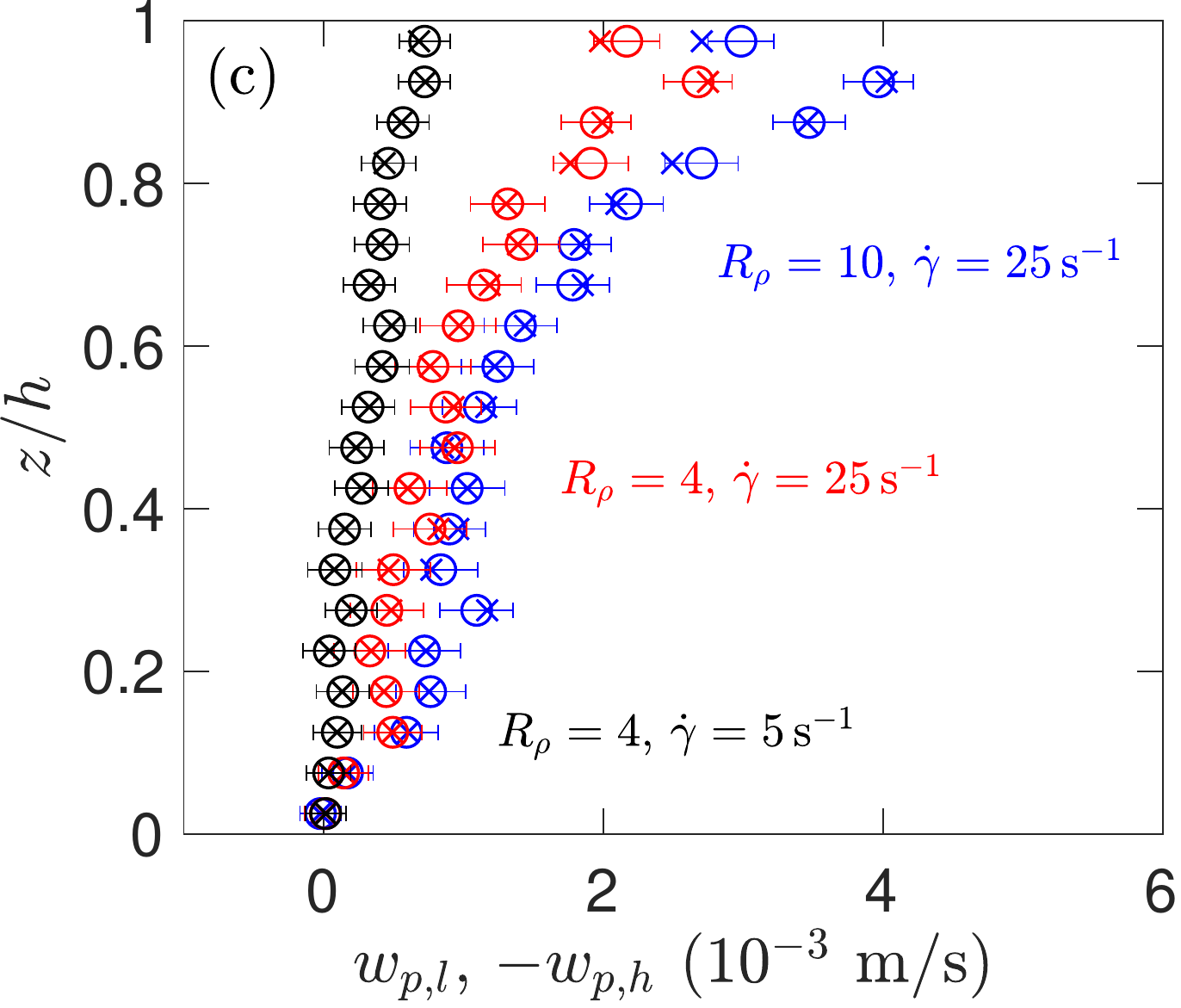}
    \subcaption{}\label{wp-i}
    \end{subfigure}
            \begin{subfigure}{0.45\columnwidth}
    \includegraphics[scale=0.494]{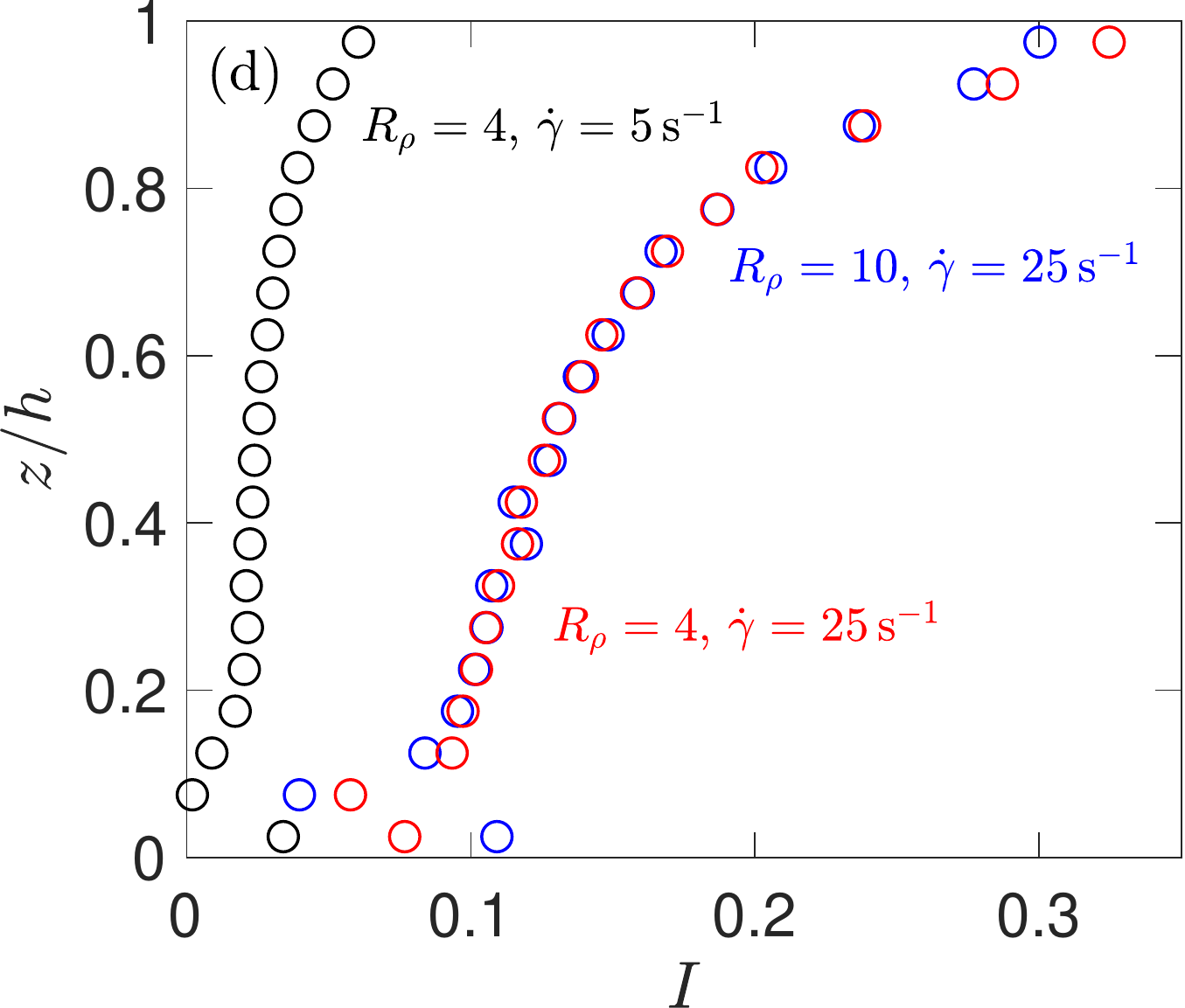}
    \subcaption{}\label{iner}
    \end{subfigure}

    \caption{ Determining the segregation velocity. (a) Average center of mass displacement $\Delta\bar{z}/d$ time series at different vertical positions under uniform shear ($\dot \gamma=U/h=25~ {\text{s}^{-1}}$, $R_\rho=8$, $c_h=c_l=0.5$, $\mu=0.2$, and $e=0.9$). (b) The heavy particle concentration profile remains relatively constant $1~$s after shear onset for the flow in Fig.~\ref{linear}. (c) Segregation velocity profiles for light particles $w_{p,l}$ ($\textcolor{black}{\times}$) and heavy particles $w_{p,h}$ ($\textcolor{black}{\Circle}$) are calculated as the average rate of change of $\Delta \bar z$ over the first $1\,$s of the simulation for flows with different density ratios and shear rates ($c_h=c_l=0.5$, $\mu=0.2$, and $e=0.9$). Both shear rate $\dot\gamma$ and density ratio $R_\rho$ affect the segregation velocity profile. Error bars represent the uncertainty in determining $\Delta\bar{z}$ and $w_{p,h}$. (d) Inertial number profiles for the flows in Fig.~\ref{wp-i}. The local inertial number is estimated as $I=\dot \gamma d/ \sqrt{ {\phi_{solid} g(1.05h-z)}}$ by substituting the local overburden pressure $P=P_{wall}+\rho_{solid}\phi_{solid}g(h-z)$ into Eq.~(\ref{inert}).
}
    \label{figg3} 
\end{figure}

In the DEM simulations, a density bidisperse mixture of $d=4~$mm spherical particles with collision time $t_c=1.25\times 10^{-4}\,$s is sheared between top and bottom horizontal frictional planes in a domain with periodic boundary conditions in the streamwise ($x$) and spanwise ($y$) directions as shown in Fig.~\ref{scheme}.  The massive planar top wall moves horizontally with constant velocity $U$ and is free to move vertically. The position of the top wall is determined by the wall weight and contact forces from the top layer of particles. The distance between the top and bottom walls $h$ remains relatively constant (fluctuating by $\pm 2$\%) after an initial rapid dilation of the particles at flow onset. The domain extends about $70d$, or $0.28\,$m, in the streamwise direction and $10d$, or $0.04\,$m, in the spanwise direction. The weight of the top wall is based on the configuration of the system so that the overburden pressure applied by the wall to the system is $P_{wall}=0.05\rho_{solid}\phi_{solid}gh$ for all cases in this study. There are 36864 particles in the system, differentiated by their densities ($\rho_h$ for heavy particles and $\rho_l$ for light particles having density ratio $R_\rho=\rho_h/\rho_l$).
The initial dense packing is achieved by placing particles in a grid pattern and letting them settle under gravity. 
The depth of the flow $h$ in the vertical $z$-direction is approximately $50d$ or $0.2\,$m,  and the volume fraction ratio ($\phi_h/\phi_l$) or, equivalently, the concentration ratio ($c_h/c_l$), is approximately uniform across the domain. Varying $c_h/c_l$ or $\rho_h/\rho_l$ in the simulations causes the segregation velocity to vary.
Over 200 simulations are performed over the wide range of flow conditions and particle properties listed in Table~\ref{table1}.

To achieve specific shear rate profiles, the corresponding velocity profile is imposed on the particles by applying a streamwise stabilizing force on each particle $k$ at every time step according to
\begin{equation}
    F_{stabilize,k}=K_s [u(z_k) - u_k],
    \label{stable}
\end{equation}
where $u(z)$ is the imposed velocity profile, $u_k$ is the particle's streamwise velocity, $z_k$ is the particle's vertical position, and $K_s$ is a gain parameter.
Details of the approach for the imposed velocity profile are provided elsewhere \cite{fry2018effect}. 
Three different profiles are considered: $u=Uz/h$, $Ue^{2.3(z/h-1)}$, and $Uz^2/h^{2}$.
The first and second profiles correspond to those for uniform shear \cite{fry2018effect} and free surface flow down a heap \cite{fan2013kinematics}, respectively.
The third profile reflects a situation that sometimes occurs in shear flows where the shear rate vanishes at a rough bottom wall \cite{jing2017effect}.

The simulation domain is divided into 20 horizontal layers for averaging purposes; each layer is $2.5d$ in the $z$-direction.
The streamwise mean velocity profile for a density segregation simulation $0.1\,$s after flow onset is shown in Fig.~\ref{uprofile} for each of the three imposed velocity profiles.
Due to the stabilizing force, the mean velocity matches the imposed profile, with a shear rate that is either uniform ($\dot\gamma=U/h$) or decreases through the depth of the particle bed $[\dot\gamma=2Uz/h^2$, $2.3Ue^{2.3(z/h-1)}/h]$.
The different symbols in Fig.~\ref{uprofile} represent heavy and light particles, which have similar streamwise velocities.
Slight deviations from the imposed velocity profiles occur within 5$d$ of the top and bottom boundaries due to particle ordering adjacent to the flat walls \cite{jing2017effect,van2018breaking}. 
Here, we focus on the flow away from the walls to avoid artifacts related to these bounding walls.
Companion simulations with bumpy walls (particles attached randomly on flat walls \cite{jing2016characterization}) show little difference with simulations using flat walls for the particles between $0.3\le z/h \le0.7$.

To characterize the evolution of the segregation, we measure the average center of mass height for each species relative to the mean height of all particles, which is calculated as
 \begin{equation}
	\bar z_{i}=\frac{1}{N_{i}} \sum _{k \in i}^{N_{i}} {z_k}- \frac{1}{N}\sum_{k=1}^{N}z_k,
\end{equation}
where $N_i$ and $N$ are the number of particles of species $i$ and the total number of particles in the horizontal averaging layer, respectively. 
Figure \ref{linear} shows the offset of the center of mass from its initial position for both heavy and light particles in horizontal layers at three vertical locations, $z/h=0.3$, $0.5$, and $0.7$, for a uniform shear flow ($u=Uz/h$). 
The offset is represented by the segregation distance $\Delta \bar z_i=\bar z_{i} -\bar z_{i,0}$, where $\bar z_{i,0}$ is the initial center of mass. 
As the segregation progresses from shear onset, heavy particles move downward while light particles move upward. Due to solid volume conservation, $\Delta \bar z_{l}+\Delta \bar z_{h}=0$. 
Since the shear rate is uniform across the domain for the simulation data shown in Fig.~\ref{figg3}, the different segregation rates in the three different layers are a consequence of the overburden, or lithostatic, pressure \cite{fry2018effect}.
A higher overburden pressure, which corresponds here to deeper bed depth (small $z/h$), reduces the segregation. 
Different segregation rates at different vertical positions are also found for the flows with the two nonlinear velocity profiles. However, in these cases, the depth-varying segregation rate results from a combination of shear rate and overburden pressure, since the shear rate now also varies with depth.

Confined granular mixtures under shear (Fig.~\ref{scheme}) initially segregate rapidly with a nearly constant segregation velocity \cite{fry2018effect}. This is because for a short period after the flow is initiated, the local particle concentration has not yet changed enough to affect the segregation, resulting in a relatively high segregation velocity, while at later times the particles reach a steady segregated state where the segregation velocity is negligible.
We focus on the initial rapidly-segregating transient. 
The initial segregation velocity is slow enough that the $1\,$s sampling window used here falls within the interval of rapid segregation ($\approx10\,$s). 
The linear profile of the segregation distance versus time in Fig.~\ref{linear} indicates the minimal effect of concentration change on the segregation velocity.
Except for some segregation that occurs in the top layer of particles, the local concentration of heavy particles remains at its initial value of $0.5\pm0.05$ during the sampling window as shown in Fig.~\ref{concentration}. 
Note that after 1~s of shear there are no strong gradients in the concentration, indicating that diffusive fluxes are unimportant, which is assumed in this approach.
The segregation velocity of each constituent is calculated from the slope of data like that in Fig.~\ref{linear} as $w_{p,i}= (\Delta \bar z_{i,1} \pm   \epsilon_{i,1})/{\Delta t }$, where $\Delta \bar z_{i,1}$ is the offset at the end of the $\Delta t=1~$s sampling window and $\epsilon_{i,1}$ is the standard error in the calculation of $\Delta \bar z_{i,1}$, which is represented by the error bars at time $t$ slightly less than $1~$s in Fig.~\ref{linear}.

By considering many thin horizontal layers, the local segregation velocity profile can be measured, as shown in Fig.~\ref{wp-i} for three sample cases of uniform shear flow.
The segregation velocity increases with both increasing density ratio and shear rate.
For each case, the segregation velocity decreases from top to bottom even though the shear rate and density ratio do not vary with depth. 
This decrease is proportional to the local inertial number in Fig.~\ref{iner}, which decreases with depth due to increasing pressure. 
This is a consequence of the overburden pressure $P$ increasing with depth, consistent with previous results \cite{fry2018effect}.
Note that the sum of segregation velocities of heavy and light particles should equal zero due to volume conservation for $c_h=c_l=0.5$. Figure \ref{wp-i} indeed demonstrates that the segregation velocities are nearly equal (and, of course, opposite).
The slight difference between $w_{p,l}$ and $-w_{p,h}$ is likely caused by a small amount of segregation during the initial filling process before the start of the shear flow. The error bars represent uncertainties in the measurement of the slope in Fig.~\ref{linear}.  

From the momentum balance equation, Eq.~(\ref{momentum}), the convection term in the $z$-direction $\frac{\partial }{\partial z}(\rho_i \phi_i w_i^2)$, which can be estimated from the simulation results, is three orders of magnitude less than the net buoyant force $-c_i\frac{\partial P}{\partial z}-\rho_i \phi_i g$ (noting that the constituent velocity $w_i$ is equal to the segregation velocity $w_{p,i}$ since the bulk velocity in the $z$-direction is zero). Likewise, the unsteady term $\frac{\partial}{\partial t}(\rho_i \phi_i w_i)$ is also negligible. This confirms that the system is in steady state during the sampling window and that the segregation velocity results from the balance between the net buoyant force and the interspecies drag as indicated by the simplified momentum expression in Eq.~(\ref{dragmomentum}).

\subsection{Inertial number dependence}

\begin{figure}
   \captionsetup[subfigure]{labelformat=empty}
    \captionsetup{justification=raggedright}
   \captionsetup[subfigure]{labelformat=empty}
    \captionsetup{justification=raggedright}
    \begin{subfigure}{0.37\columnwidth}
    \includegraphics[scale=0.48]{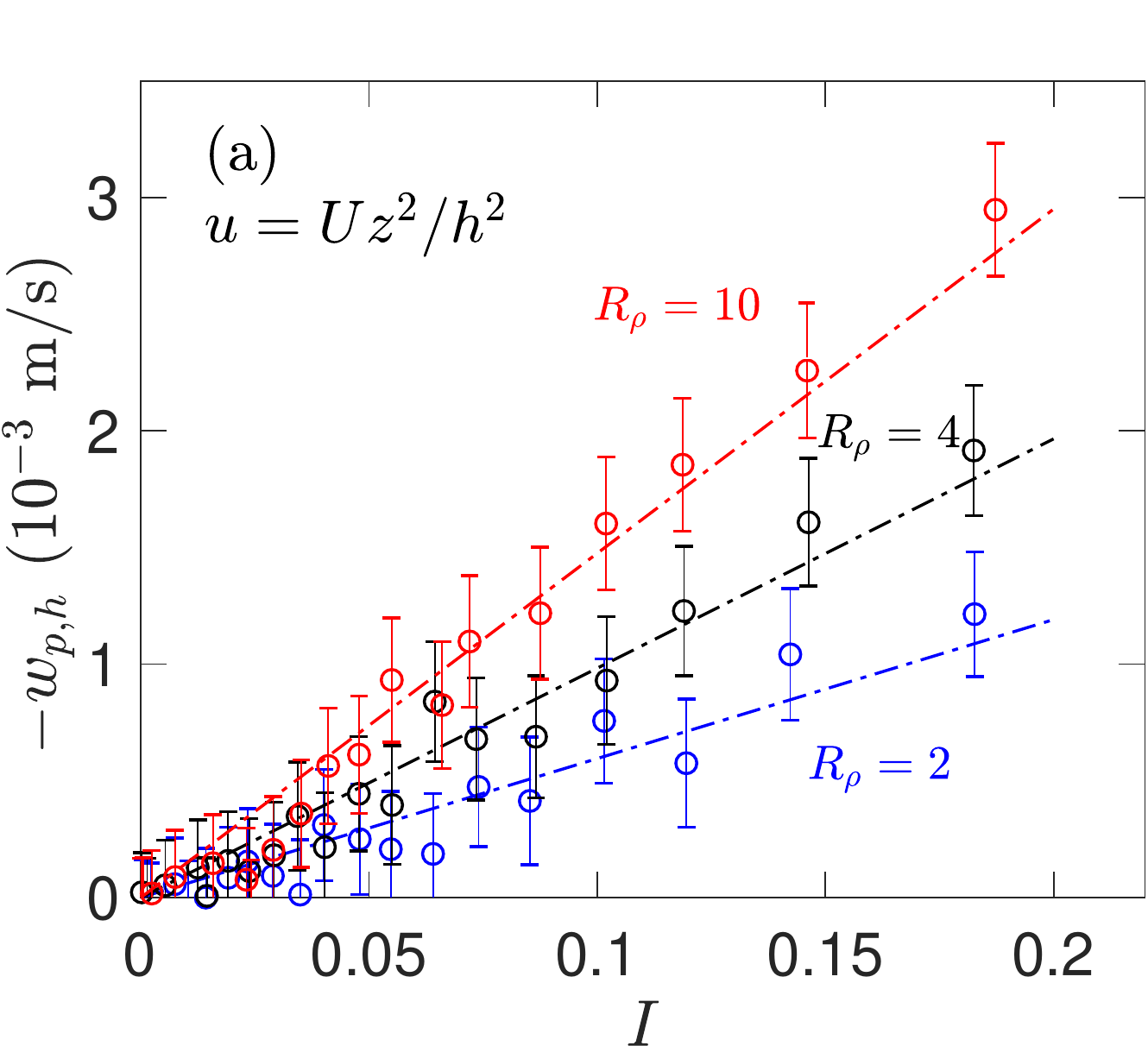}
    \subcaption{}\label{}
     \end{subfigure}
    \begin{subfigure}{0.37\columnwidth}
    \includegraphics[scale=0.48]{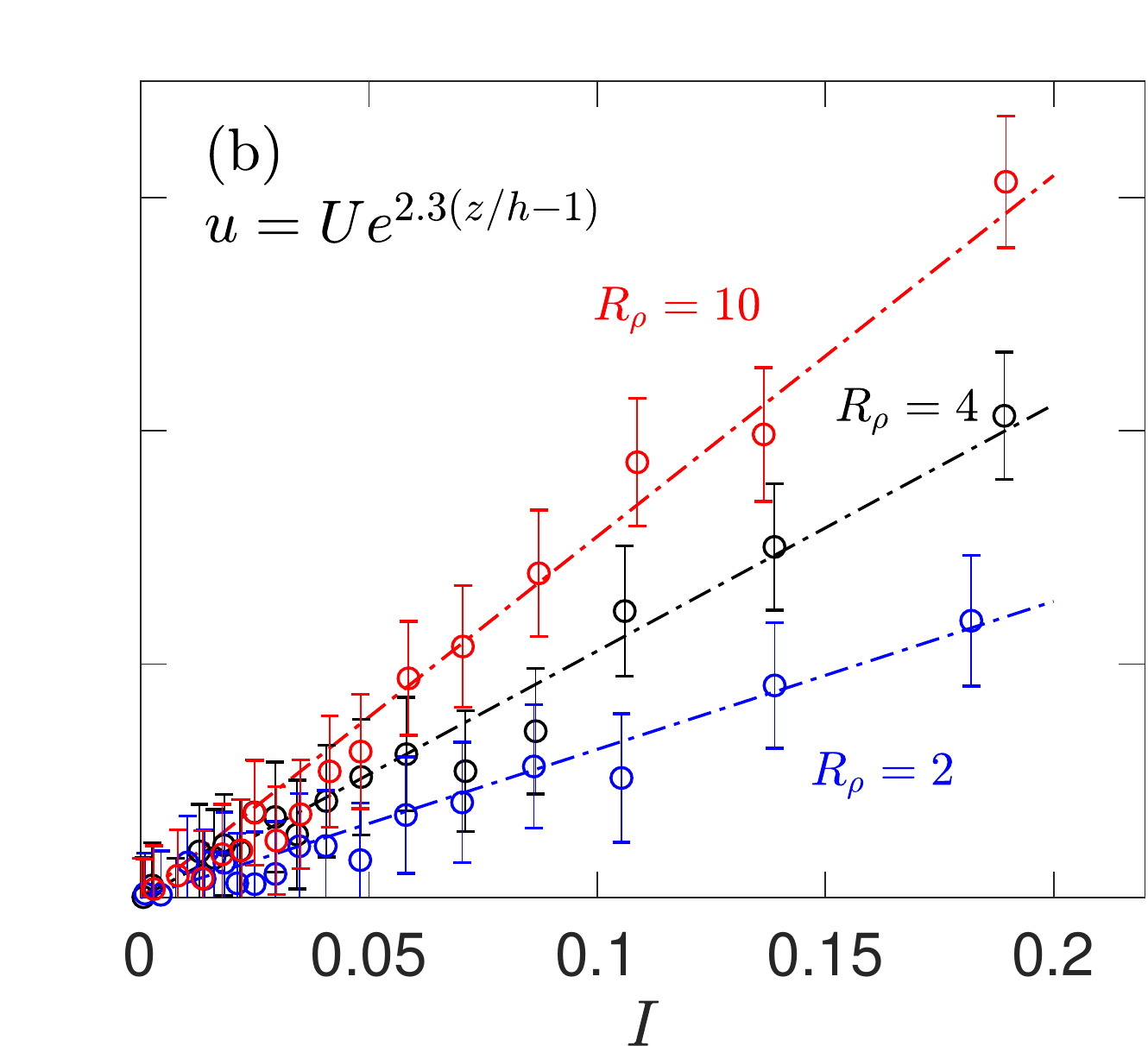}
    \subcaption{}\label{}
        \end{subfigure}
     \caption{Segregation velocities of light particles vs.\ local inertial number $I$ at different depths for nonlinear velocity profiles (a) $u=U z^2/h^2$ and (b) $u=Ue^{2.3(z/h-1)}$ for simulations with $U=2~$m/s, $R_\rho=\rho_h/\rho_l \in \{ 2, 4, 10 \} $, $c_h=c_l=0.5$, $\mu=0.2$, and $e=0.9$. Dashed lines are linear fits for each density ratio, demonstrating a linear dependence of $w_{p,l}$ on $I$. Error bars represent the uncertainty in determining $w_{p,l}$ as indicated in Fig.~\ref{figg3}.}
         \label{wp-nonlinear}
\end{figure}

\begin{figure}
   \captionsetup[subfigure]{labelformat=empty}
    \captionsetup{justification=raggedright}
    \begin{subfigure}{0.37\columnwidth}
        \includegraphics[scale=0.48]{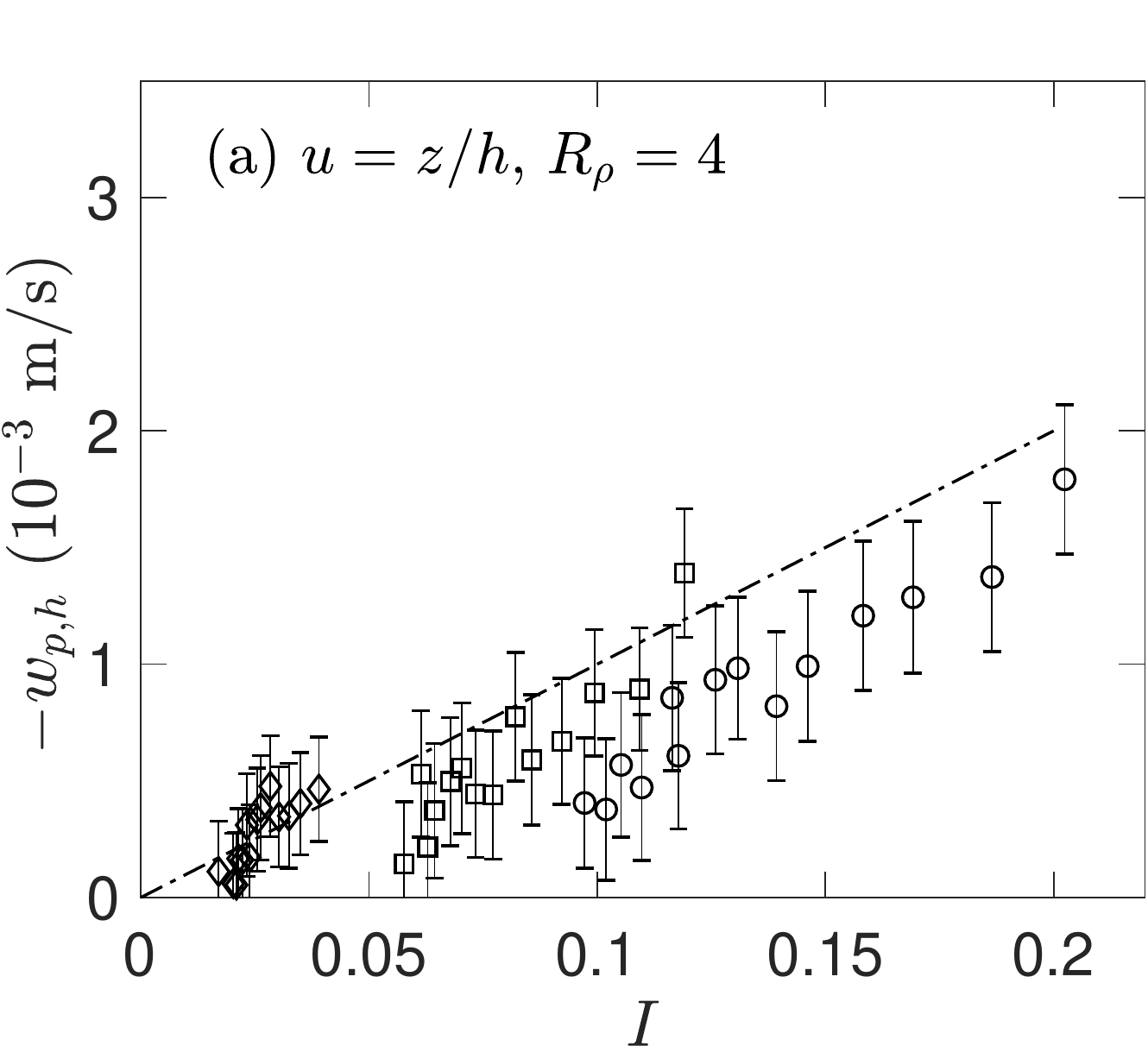}
    \subcaption{}\label{wp-shear}
     \end{subfigure}
    \begin{subfigure}{0.37\columnwidth}
       \includegraphics[scale=0.48]{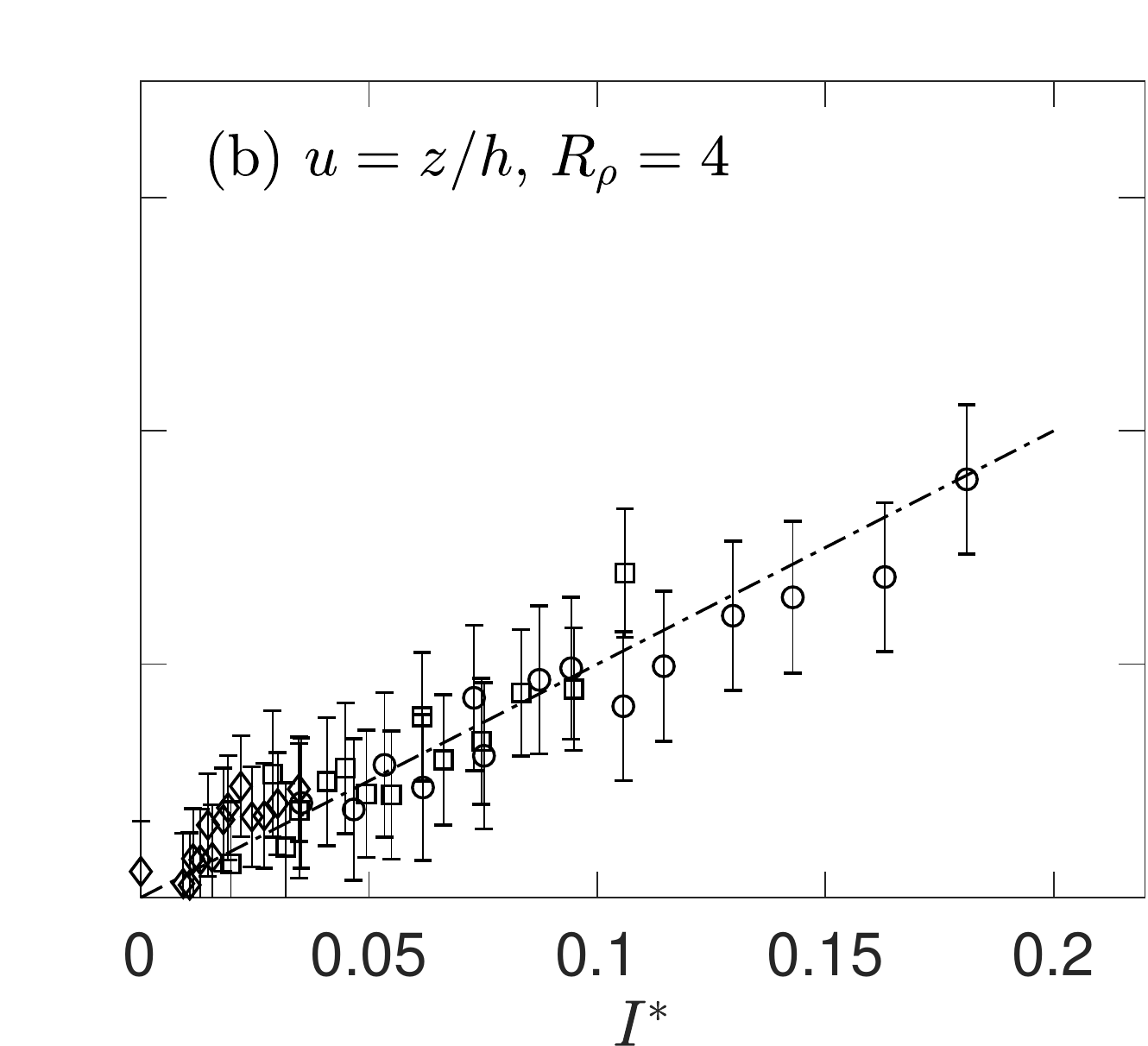}
    \subcaption{}\label{wp-gamma}
    \end{subfigure}
    \caption{Segregation velocities of light particles $w_{p,l}$ for $0.2\le z/h \le0.8$ vs.\ (a) inertial number $I$ and (b) modified inertial number $I^*=\sqrt{I^2-I_0^2}$  for simulations with particle properties given in Fig.~\ref{wp-nonlinear} but for uniform shear rate profiles with $\dot \gamma=$  $5~\text{s}^{-1}$ (\textcolor{black}{$\Circle$}), $15~\text{s}^{-1}$ (\textcolor{black}{$\Square$}), and $25~\text{s}^{-1}$ (\textcolor{black}{$\Diamond$}). $I_0$ is the cut-off inertial number at which density segregation effectively ceases \cite{fry2018effect}. The dashed line in both figures is identical and a linear fit to the data in (b). Error bars represent the uncertainty in determining $w_{p,l}$ as indicated in Fig.~\ref{wp-i}. }    
    \label{wp-linear}
\end{figure}

The inertial number dependence in the proposed drag model, Eq.~(\ref{eq8}), requires that the segregation velocity is linear in $I$ [Eqs.~(\ref{final}) and (\ref{finalh})]. 
Therefore, to verify the drag model, segregation velocities of light particles $w_{p,l}$ at different depths are plotted versus $I$ for different density ratios $R_\rho$ in Fig.~\ref{wp-nonlinear} for flows with nonlinear streamwise velocity profiles. 
Similar results are also found for the segregation velocity of heavy particles $w_{p,h}$ since $w_{p,h}\approx w_{p,l}$ for flows with $c_h=c_l=0.5$.
The linear relation between $w_{p,i}$ and $I$ in Fig.~\ref{wp-nonlinear} for varying $\dot\gamma$ and $P$ (expressed in terms of varying $I$) confirms the assumed dependence on $I$ and is consistent with results from previous studies \cite{liu2017transport,fry2018effect}.
That is, the segregation velocity depends linearly on $I$ for each density ratio. The error bars are somewhat large, consistent with the difficulty in accurately measuring $w_{p,i}$ as indicated in Fig.~\ref{figg3}. Nevertheless, a line can be fit through the data at each density ratio.
Furthermore, the fitting lines are quite similar for both nonlinear velocity profiles.
 
The segregation velocity data, however, do not show as clear a linear dependence on $I$ when plotted in the same way for uniform shear flows, as illustrated in Fig.~\ref{wp-shear}. This is because unlike the other two nonlinear velocity profiles $[ u=U z^2/h^2$ and $Ue^{2.3(z/h-1)} ]$, the local inertial number in a uniform shear flow is not zero or close to zero at the bottom wall.
Instead, in the case of the linear velocity profile ($u=Uz/h$), segregation is forced to cease at the fixed bottom wall because of the wall itself.
To account for the cessation of density segregation at the bottom wall, we use a modified inertial number for uniform shear flows
\begin{equation}
	I^*=\sqrt{I^2-I_0^2},
	\label{istar}
\end{equation}
where $I_0$ is the inertial number at which density segregation effectively ceases.
According to Fig.~\ref{wp-i}, segregation ceases at $z/h\approx0$. Therefore, we assume $I_0$ is the inertial number at that location, such that $I_0=\dot \gamma d \sqrt{ {\rho_{bulk}}/{P_{0}}}$ with $P_0=\rho_{bulk}gh+ P_{wall}$.
Modifying $I$ with $I_o$ in Eq.~(\ref{istar}) forces $w_{p,i}$ to 0 at the bottom wall and brings the nonlocal effect of boundaries into the proposed model.
An alternative approach for considering the boundary is to solve a second state variable related to the velocity fluctuations (i.e., ``granular temperature" in KTGF \cite{jenkins1983theory} or ``granular fluidity" in the nonlocal rheology of Ref. \cite{kamrin2012nonlocal}) from an independent equation supplied with boundary conditions such that the boundary effect is considered implicitly in the drag. 
However, rheology models such as $\mu(I)$ and KTGF have different coefficients for different particle properties, which requires simulation data to find those coefficients first. Furthermore, the boundary condition that determines the ``heat flux" (i.e., particle velocity fluctuations induced by boundaries) into the system often involves fitting parameters \cite{johnson1987frictional}, which makes it harder to generalize the model. As such, we do not consider velocity fluctuations here. Instead, the modified inertial number is used to correct the segregation near the boundary.
As shown in Fig.~\ref{wp-gamma}, using $I^*$ instead of $I$ collapses the data for uniform shear flows such that the dependence of $w_{p,i}$ on $I^*$ is close to linear. 
Note that for the two nonlinear velocity profiles $I^*$ equals $I$ since $I_0 \approx 0$. 
From here on, unless otherwise specified, we drop the asterisk on $I$ and use the corrected inertial number [Eq.~(\ref{istar})].

Figures \ref{wp-nonlinear} and \ref{wp-gamma} show that $w_{p,i}$ depends linearly on $I$ for varying shear profiles, shear rates, and density ratios. This confirms  $\beta_i \propto I^{-2}$ in Eq.~(\ref{eq8}) since equating the buoyant and drag forces gives $w_{p,i}\propto I$ in Eqs.~(\ref{final}) and (\ref{finalh}). 
The linear relation between $w_{p,i}$ and $I$ is also consistent with previous results in free surface flows. Consider, for example, heap flows, in which the flowing layer thickness is about eight particle diameters \cite{fan2013kinematics,xiao2016modelling}. For such a thin layer with the local shear rate decreasing exponentially with depth, the overburden, or lithostatic, pressure can be treated as a uniform parameter, and the linear relation between $w_{p,i}$ and $I$ can be reduced to a linear relation between $w_{p,i}$ and $\dot\gamma d$ \cite{fan2014modelling,schlick2015modeling}.
On the other hand, for a thick flowing layer with a uniform shear rate, the segregation velocity at different depths in the flowing layer is only affected by the overburden pressure so long as we consider segregation away from the upper and lower bounding walls. The effect of lithostatic pressure has to be included, resulting in a dependence of $w_{p,i}$ on $I$ \cite{liu2017transport,fry2018effect}.
We further note that in our previous work \cite{fry2018effect}, we proposed a form for $I^*$ in terms of a critical pressure that is equivalent to Eq.~(\ref{istar}).
However, it should be noted that at larger inertial numbers ($I>0.5$) \cite{liu2017transport}, instantaneous binary collisions dominate over enduring contacts, resulting in a rheological change from dense to rapid dilute granular flow. In the latter regime, the linear relation between $w_{p,i}$ and $I$ is likely to be invalid.

\subsection{Density ratio dependence}

\begin{figure}
    \captionsetup{justification=raggedright}
    \centerline{\includegraphics[scale=0.5]{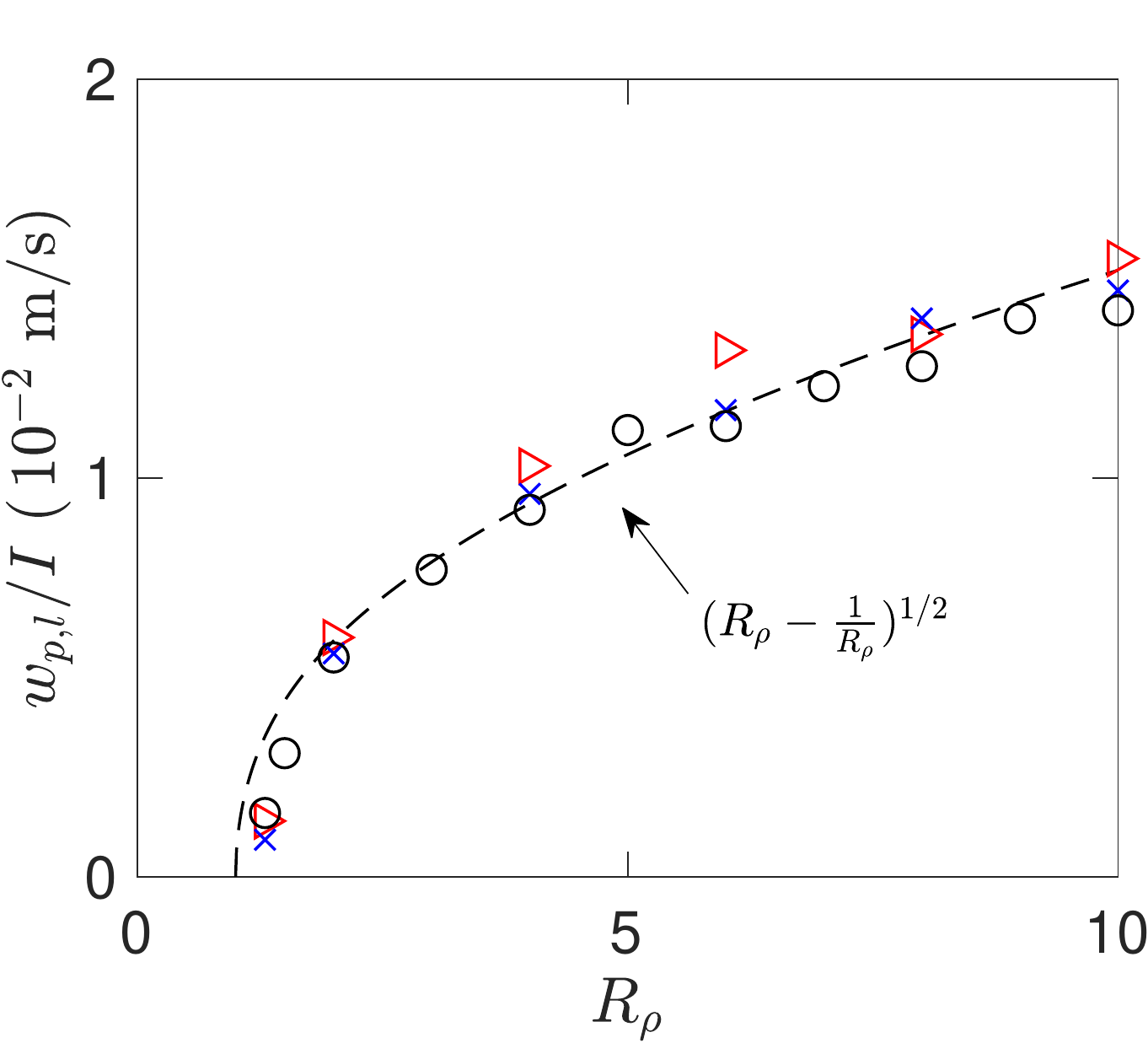}}
    \caption{${w_{p,l}/I}$ vs.\ $R_\rho$ for simulations with $1.3\le R_\rho \le10$, $\mu=0.2$, $e=0.9$, and $c_h=c_l=0.5$. Colors and symbols represent simulations with $\dot\gamma=U/h$ ($\textcolor{black}{\Circle}$), $2Uz/h^{2}$  ($\textcolor{blue}{\times}$), and $2.3Ue^{2.3(z/h-1)}/h$ ($\textcolor{red}{\rhd}$). Dashed curve shows the dependence of $w_{p,i}/I$ on $R_\rho$ from Eq.~(\ref{final}) with $B(\mu)=700$.}
    \label{denwpi}
\end{figure}

As shown in Fig.~\ref{wp-nonlinear}, $w_{p}/I$ increases with density ratio $R_\rho$.
Equations~(\ref{final}) and (\ref{finalh}) indicate that $w_{p}/I$ is proportional to $\sqrt{R_\rho-1/R_\rho}$.
To demonstrate this dependence, $w_{p,l}/I$, calculated as the slope of the fitting lines in Figs.~\ref{wp-nonlinear} and \ref{wp-gamma}, is plotted versus $R_\rho$ in Fig.~\ref{denwpi}.
The dashed curve in Fig.~\ref{denwpi} represents the predictions of Eq.~(\ref{final}) with $B(\mu=0.2)=700$. 
(We characterize the dependence of $B$ on $\mu$ in Section \ref{section3}.D.)
Indeed, $w_{p,l}/I$ is proportional to $\sqrt{R_\rho-1/R_\rho}$ over the wide range of $R_\rho$ tested ($1.3\le R_\rho \le 10$) in different flow profiles where $I$ varies due to changes in both $P$ and $\dot\gamma$. 
Agreement between the curve and the data confirms the dependence of the drag model on particle densities and, equivalently, the functional dependence of $w_{p,i}$ on $R_\rho$ in Eqs.~(\ref{final}) and (\ref{finalh}).


\subsection{Concentration ratio dependence}

\begin{figure}
    \captionsetup{justification=raggedright}
    \centerline{\includegraphics[scale=0.55]{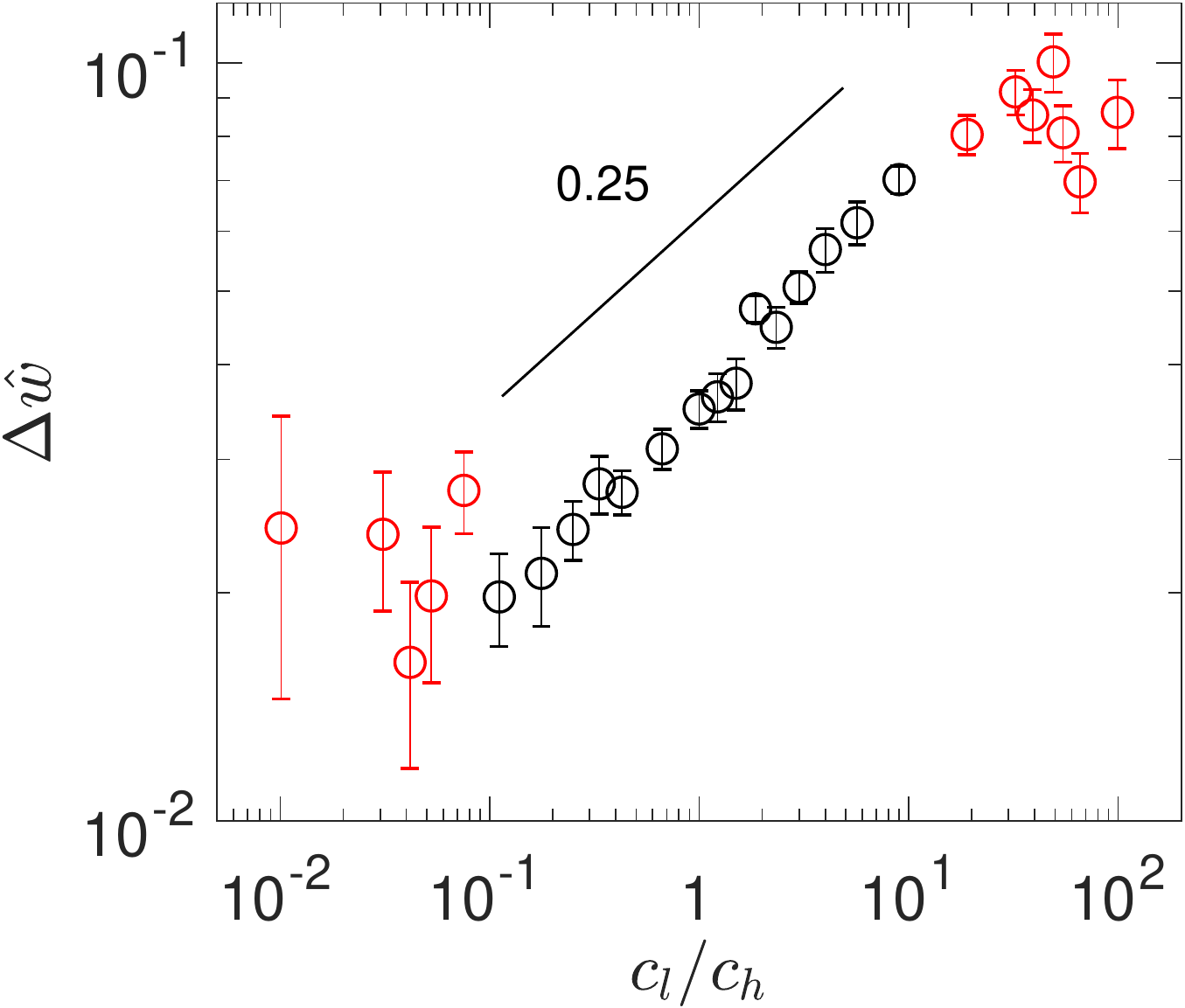}}
    \caption{Normalized velocity difference between species $\Delta \hat{w}$ vs. concentration ratio ${c_l/c_h}$ for simulations with $R_\rho=4$, $\mu=0.2$, $e=0.9$, and $\dot\gamma=U/h=25~\text{s}^{-1}$. Error bars represent uncertainty in determining $\Delta \hat{w}$. Strong correlation between $\Delta \hat{w}$ and $c_l/c_h$ is found for $0.1 \leq c_l,c_h \leq 0.9$ ($\Circle$), as opposed to situations approaching a single intruder particle at concentrations outside this range, $c_l$, $c_h \leq 0.1$ or $c_h$, $c_l \geq 0.9$ (\textcolor{red}{\Circle}). Solid line with slope 0.25 is consistent with the slope of the data, confirming the dependence of the drag model on particle concentrations for $0.1 \leq c_l,c_h \leq 0.9$.}
    \label{wpcon}
\end{figure}

A recent study \cite{jones2018asymmetric} indicates that segregation in mixtures of heavy and light particles has an underlying asymmetry that depends on local particle concentration, similar to mixtures where particles differ in size \cite{van2015underlying,jones2018asymmetric}. Specifically, a heavy particle among mostly light particles segregates faster than a light particle among mostly heavy particles. 
The last term $\sqrt{\phi_h/\phi_l}$, on the right hand side of the drag model in Eq.~(\ref{eq8}) accounts for the nonlinear dependence of segregation velocity on particle concentration.
To confirm this term, we rewrite Eq.~(\ref{final}) using $w_{p,l}=c_h( w_l - w_h)$ from Eq.~(\ref{bulkv}), such that $c_h$ on the l.h.s and $(1-c_l)$ on the r.h.s cancel and the equation becomes
\begin{equation}
	{(w_l - w_h)^2} =\frac{1}{B(\mu)}\frac{gd}{ \phi_{solid}}  (R_\rho-\frac{1}{R_\rho}) {I^2} \sqrt{\frac{c_l}{c_h}}  . 
	\label{kk}
\end{equation}
According to Eq.~(\ref{kk}), $w_l - w_h$, based on the proposed drag model of Eq.~(\ref{eq8}), should depend linearly on $({c_l}/{c_h})^{1/4}$.
To test this dependence, we keep particle densities and other simulation parameters constant, but vary $c_h$ from 0.01 to 0.99 for uniform shear flow.
Figure \ref{wpcon} plots the normalized velocity difference between the two species $\Delta \hat{w}=\frac{w_l-w_h}{I} \sqrt{ \frac{\phi_{solid}}{gd} (R_\rho-\frac{1}{R_\rho})^{-1} }$, excluding $B(\mu)$ because $\mu$, and hence $B(\mu)$, is the same for all simulations included in the figure, as a function of ${c_l/c_h}$ for uniform shear flows. 
Note that $(w_{p,l}-w_{p,h})/I$ is calculated using the slope of the fitting line through the data at different depths like that in Fig.~\ref{wp-gamma}.
This avoids propagating the random error evident in these figures as the deviation of the data points for individual simulations from the fitting line.
From Fig.~\ref{wpcon} it is evident that $\Delta \hat{w}$ depends linearly on $({{c_l}/{c_h}})^{1/4}$ for $c_i$ in the range 0.1 to 0.9.
Outside of this range the number of particles of each species decreases to a point where the segregation velocity of the low concentration species is difficult to accurately determine. As a result, the uncertainty of $\Delta \hat{w}$ is large for ${{c_l}/{c_h}}<0.1$.
Apart from the large uncertainty, the relative velocity and the concentration are uncorrelated for $c_i<0.1$. This is because the segregation of the low concentration species is tending toward that in the single particle intruder case. For intruder particles, the concept of concentration loses its physical meaning, and the continuum assumption required by the approach in this paper is no longer valid. For example, the difference in the segregation velocities between a single intruder case and a ten-intruder case is minimal as long as the intruders are well separated by the bed particles. The ‘concentration’ increases by ten times, but the relative velocity remains constant.
Nonetheless, Fig.~\ref{wpcon} confirms the linear relation between $\Delta \hat{w}$ and $({{c_l}/{c_h}})^{1/4}$ used in Eqs.~(\ref{final}) and (\ref{finalh}), and thus the proposed form of the KTGF drag model [Eq.~(\ref{eq8})] for flows with $0.1\leq c_i \leq 0.9$.

\subsection{Friction and restitution dependence }
\label{sectionB}

\begin{figure}
   \captionsetup[subfigure]{labelformat=empty}
    \captionsetup{justification=raggedright}
    \begin{subfigure}{0.37\columnwidth}
    \includegraphics[scale=0.494]{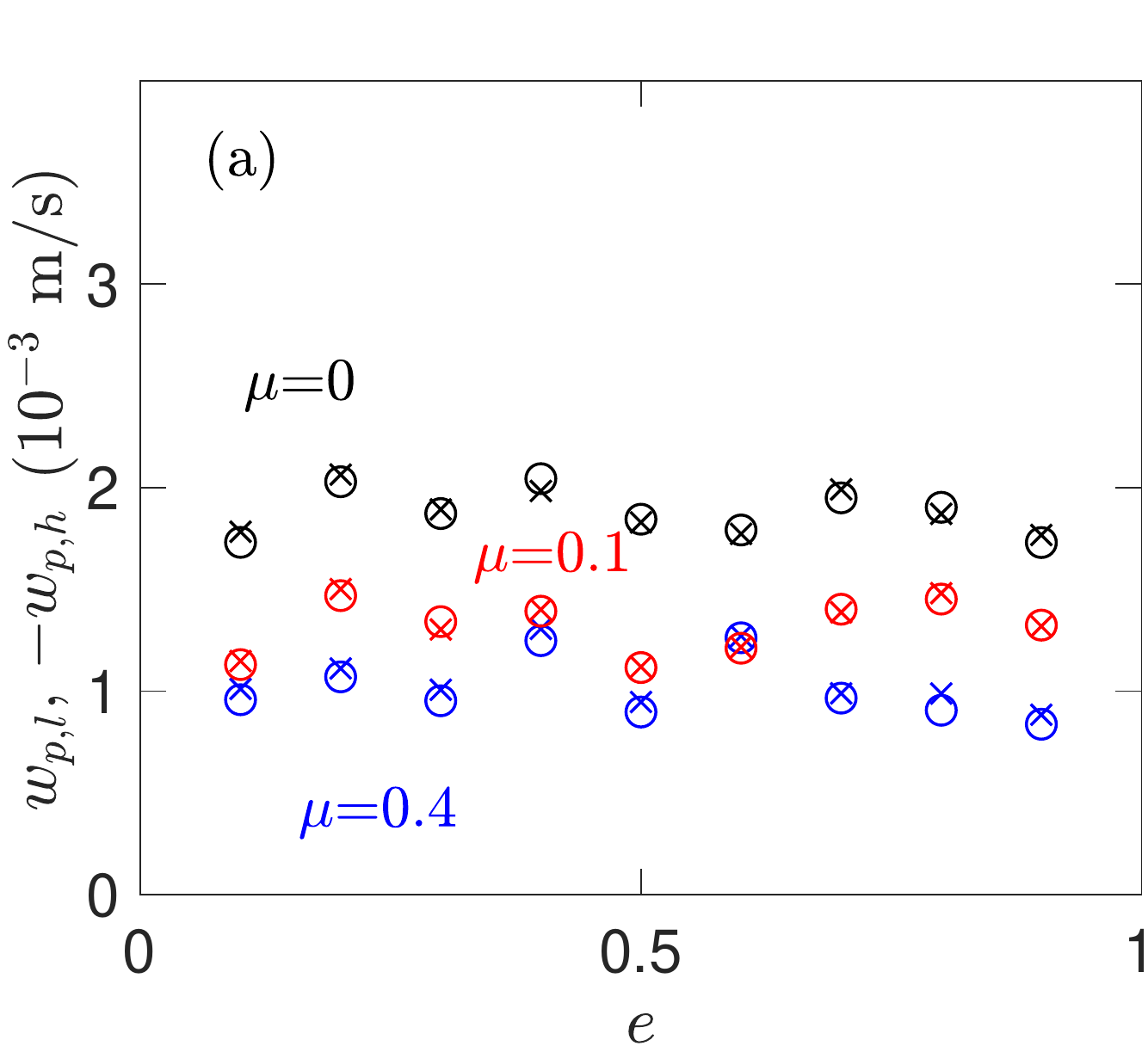}
    \subcaption{}\label{wpe}
     \end{subfigure}
    \begin{subfigure}{0.37\columnwidth}
    \includegraphics[scale=0.494]{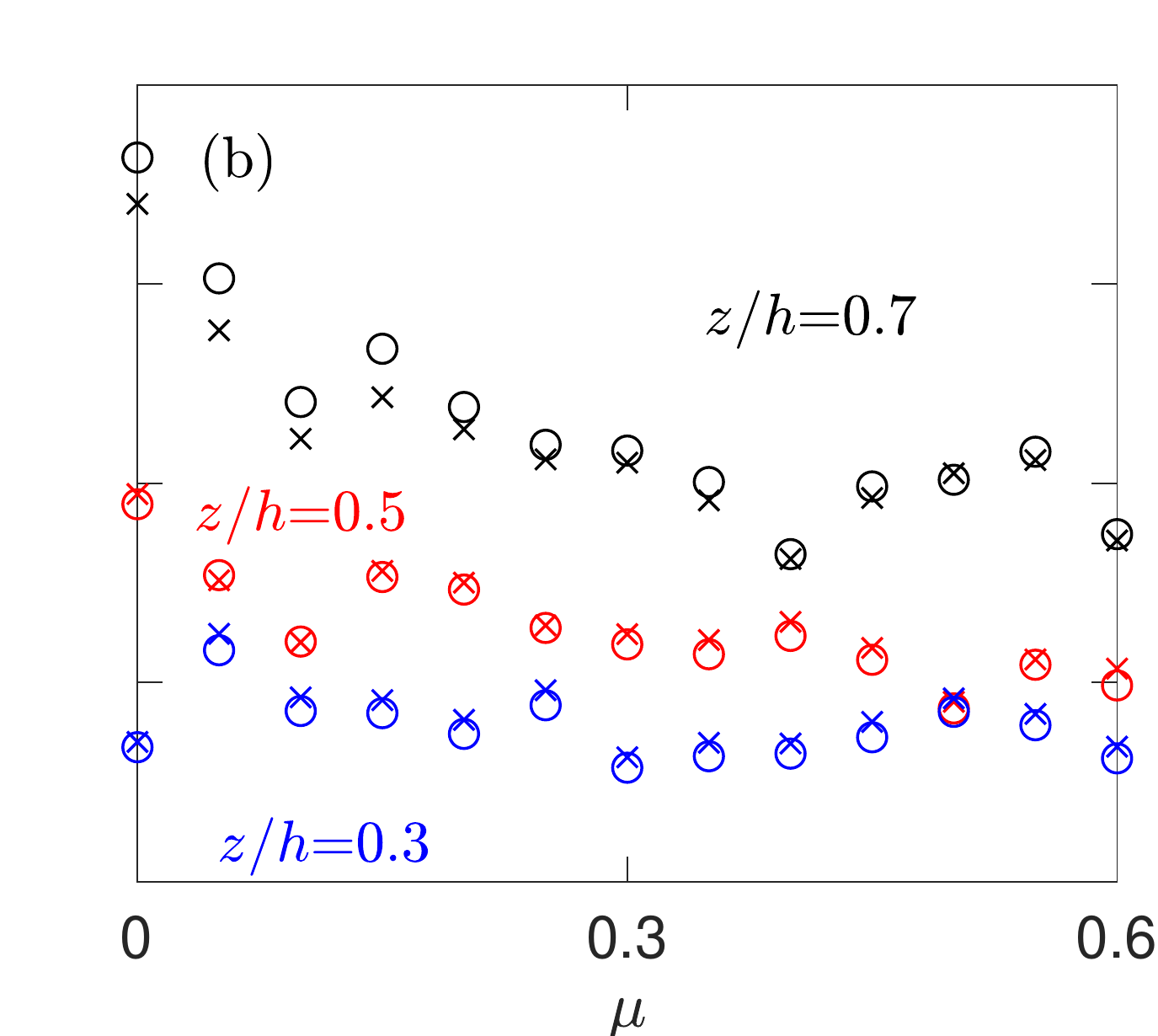}
    \subcaption{}\label{wp-mu}
    \end{subfigure}
    \caption{Dependence of segregation velocities $w_{p,i}$ on contact parameters. (a) Light ($\times$) and heavy ($\Circle$) particle segregation velocities are nearly independent of restitution coefficient $e$ at $z/h=0.5$ with $\mu=0$ (\textcolor{black}{black}), $0.1$ (\textcolor{red}{red}), and $0.4$ (\textcolor{blue}{blue}). (b) Particle segregation velocity decreases as $\mu$ increases for $z/h=0.3$ (\textcolor{red}{red}), $0.5$ (\textcolor{blue}{blue}), $0.7$ (\textcolor{black}{black}) with $e=0.9$. $R_\rho=8$, $c_h=c_l=0.5$, and $\dot\gamma=U/h=25~\text{s}^{-1}$.}
\end{figure}

The functional form for $B(\mu)$ characterizing the dependence of the drag on particle friction in Eqs.~(\ref{eq8}), (\ref{final}), and (\ref{finalh}) is difficult to derive analytically due to the complex particle interactions typical of dense granular flows. Instead, DEM simulations under a variety of flow conditions are used to empirically determine $B(\mu)$ and demonstrate that it is independent of the restitution coefficient $e$. 

Consider first $e$, which characterizes energy dissipation in short-duration collisions.
We vary $e$ from 0.1 to 0.9 for uniform shear flow while keeping all other parameters unchanged, such that the damping coefficient of the linear-spring-dashpot collision model used in the DEM simulations varies accordingly. The segregation velocities for both light and heavy particles at $z/h=0.5$ for the case with $c_h=c_l=0.5$ and $R_\rho=8$ are plotted versus $e$ in Fig.~\ref{wpe} for three different friction coefficients $\mu$. 
Although there is some scatter in the data, it is clear that $w_{p,i}$ is independent of $e$, regardless of $\mu$, as is reasonable to expect for dense granular mixtures in which short-duration collisions are unlikely to play as important of a role as in dilute flows.

For dense granular mixtures, enduring contacts dominate, and stresses are generated by long-duration sliding and rolling contacts. Unlike $e$, the surface friction coefficient $\mu$ has significant impact on $w_{p,i}$ as shown in Fig.~\ref{wp-mu}. The particle segregation velocity generally decreases as $\mu$ increases at different bed depths. 
Note that $w_{p,i}$ measured from DEM simulations is affected by not only $\mu$ but also the $\mu$ dependent initial packing, which adds more uncertainties resulting in scatter in the data. 
Nevertheless, the overall trend of decreasing segregation velocity with increasing friction remains evident.

\begin{figure}
   \captionsetup[subfigure]{labelformat=empty}
    \captionsetup{justification=raggedright}
    \begin{subfigure}{0.45\columnwidth}
    \includegraphics[scale=0.494]{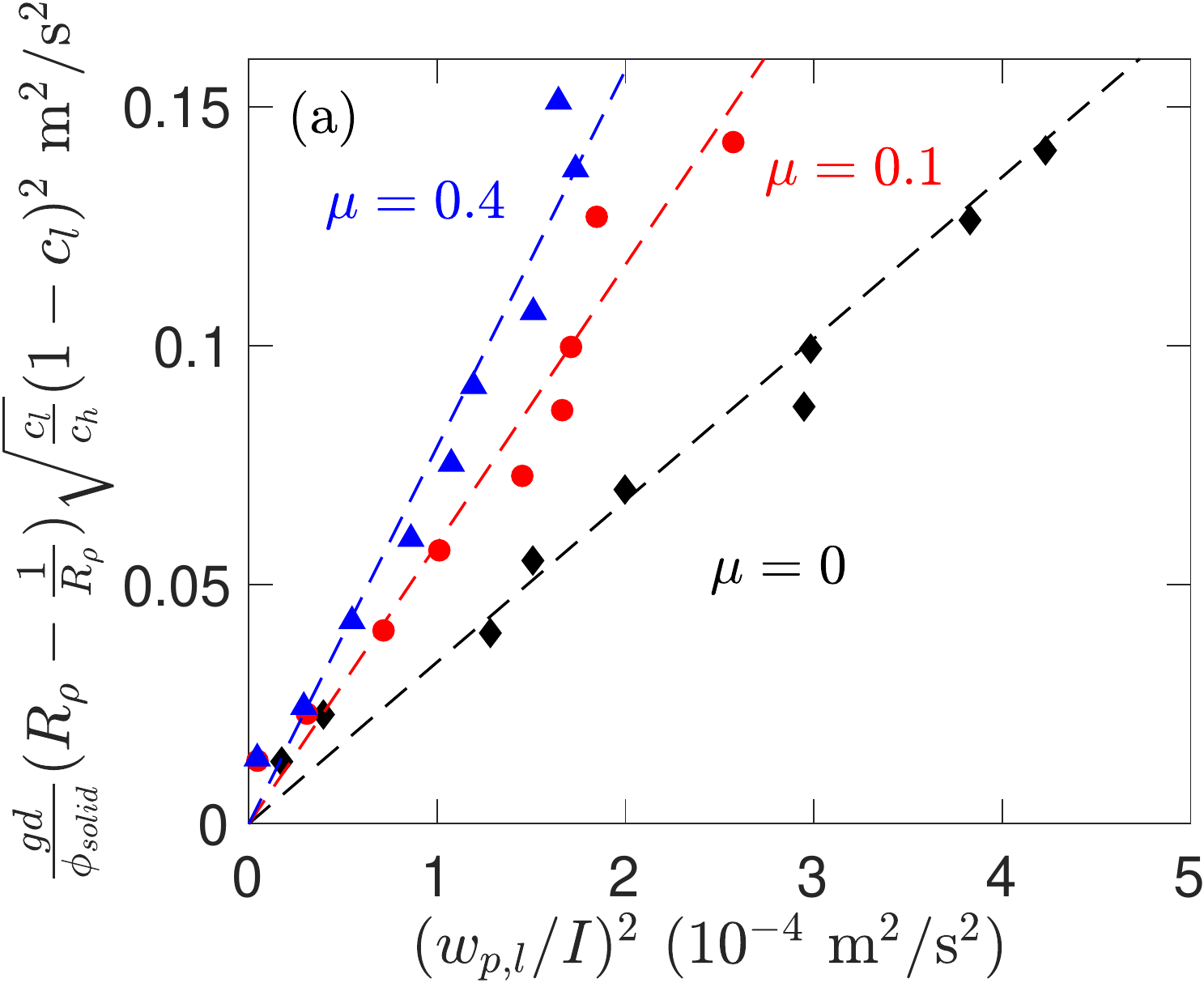}
    \subcaption{}\label{singlecoefficient}
     \end{subfigure}
    \begin{subfigure}{0.45\columnwidth}
    \includegraphics[scale=0.494]{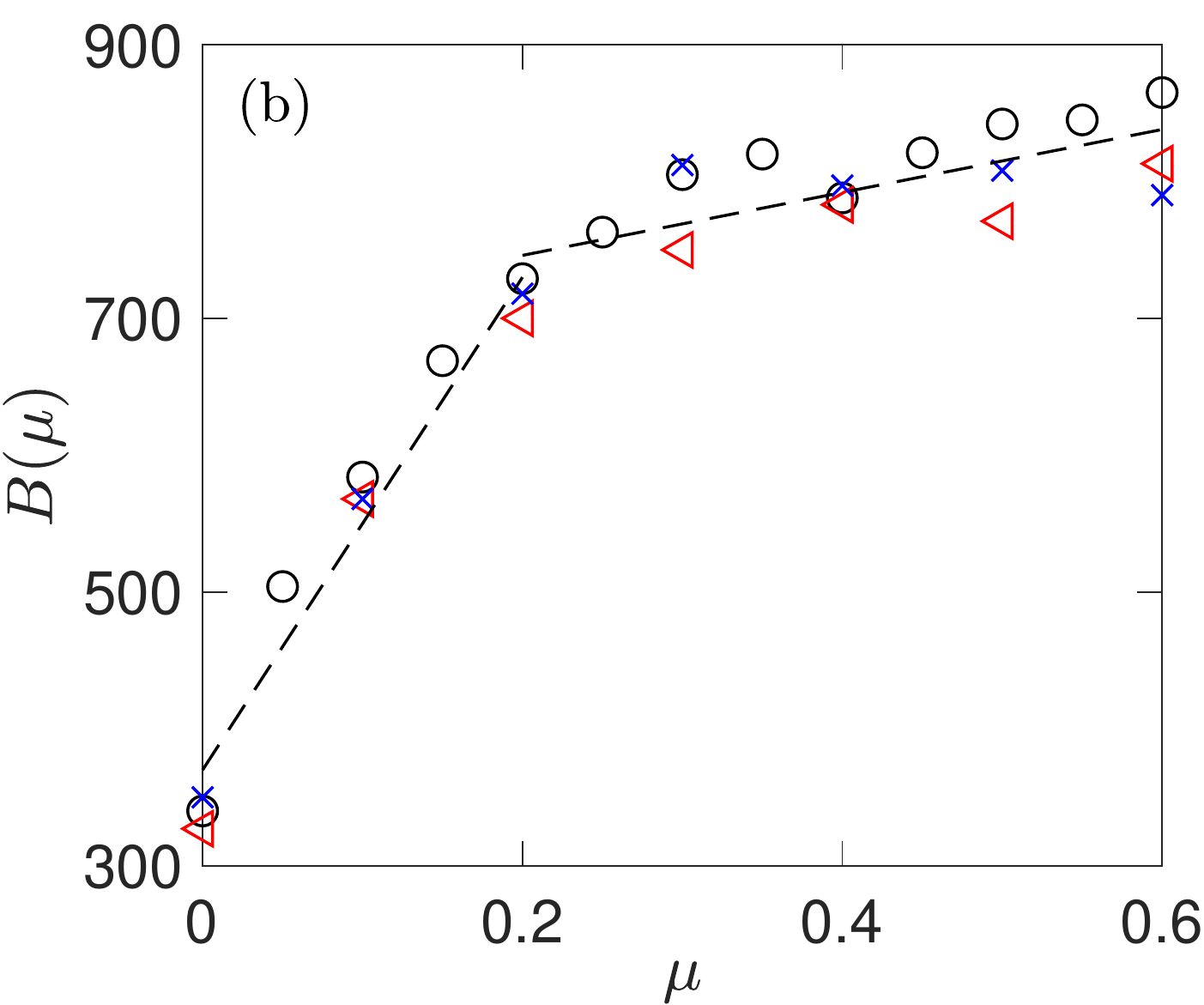}
    \subcaption{}\label{coefficientB}
    \end{subfigure}
    \caption{Determining $B(\mu)$. (a) $\frac{gd}{\phi_{solid}}(R_\rho-\frac{1}{R_\rho})\sqrt{\frac{c_l}{c_h}} (1-c_l)^2$ vs.\ $(w_{p,l}/I)^2$ for uniform shear flows with $R_\rho\in \{ 2, 4, 6, 8, 10 \}$ and three different $\mu$ values. $B(\mu)$ is determined by the fitted slope of the dashed lines, which are forced through zero ($c_h=c_l=0.5$, $\dot\gamma=25$ s$^{-1}$, and $e=0.9$). (b) $B(\mu)$ vs.\ $\mu$ based on 243 simulations. Different symbols correspond to $e=0.6$ ($\textcolor{blue}{\times}$), 0.8 ($\textcolor{red}{\lhd}$), and 0.9 ($\Circle$) for uniform shear flows. }
\end{figure}

In this context, $B(\mu)$ is determined through a series of DEM simulations for uniform shear flow with different density ratios $R_\rho$ and friction coefficients $\mu$. 
Since the dependence of drag on the other parameters has been verified, it is possible to determine the functional form for $B(\mu)$ from Eq.~(\ref{final}),

\begin{equation}
B(\mu)=\frac{gd}{\phi_{solid}}(R_\rho-\frac{1}{R_\rho})\sqrt{\frac{c_l}{c_h}} (1-c_l)^2  \Bigg/ \left(\frac{w_{p,l}}{I} \right)^2.
\label{eqc}
\end{equation}
In applying Eq.~(\ref{eqc}), the slope of the fitting line in Fig.~\ref{wp-gamma} is again used for $w_{p,l}/I$.
We estimate $B(\mu)$ by plotting $\frac{gd}{\phi_{solid}}(R_\rho-\frac{1}{R_\rho})\sqrt{\frac{c_l}{c_h}} (1-c_l)^2$ versus $(w_{p,l}/I)^2$ over a range of density ratios for three values of $\mu$ in Fig.~\ref{singlecoefficient}. The slope of the data at each $\mu$ gives the corresponding value of $B(\mu)$. 

By measuring the slopes of different sets of data with $0 \le \mu \le 0.6$, we empirically determine the dependence of $B(\mu)$ on $\mu$. The results are plotted in Fig.~\ref{coefficientB} for several values of $e$ for uniform shear flow. 
For bidisperse mixtures with a small friction coefficient ($\mu<0.3$), $B(\mu)$ is linearly proportional to $\mu$. For larger values of $\mu$ ($0.3\leq \mu \leq  0.6$), the relation for $B(\mu)$ remains linear but with a smaller slope.
There is slightly more scatter for different values of $e$ at larger $\mu$.
Note that $B(\mu)=700$ from Fig.~\ref{coefficientB} is consistent with the value used in Fig.~\ref{denwpi} to calculate $w_{p,i}/I$.

The interspecies drag is a combination of normal and tangential forces resulting from long-duration contacts. 
When $\mu=0$, the tangential forces vanish. Thus, $B(\mu)\neq 0$ means that interspecies drag of smooth particle systems is entirely due to normal forces. 
Figure \ref{coefficientB} shows that $B(\mu)$ increases with $\mu$.
That is, increasing friction increases the interspecies drag, thereby reducing the segregation velocity.
This is contrary to observations from size segregation, where part of the segregation driving force comes from interparticle friction, such that increasing friction coefficient promotes the segregation of large particles \cite{jing2017micromechanical}. A possible explanation is that the friction force adds to the segregation driving force in size segregation, which overcomes the resultant increase in the interspecies drag.
Alternately, for particles having different densities but the equal size, the friction force has no impact on the segregation driving force but adds to the drag.

Note that Eqs.~(\ref{final}) and (\ref{finalh}) are not limited to specific velocity profiles even though $B(\mu)$ is empirically determined using the data from uniform shear flows with $c_h=c_l=0.5$. $B(\mu)$ is merely a coefficient to account for the particle frictional properties. Thus, the KTGF drag model of Eqs.~(\ref{final}) and (\ref{finalh}) is generally applicable to predict density segregation in any dense segregating granular flow.%

\section{Parameters for the viscous drag model}
\label{section4}

\begin{figure}
    \captionsetup{justification=raggedright}
    \centerline{\includegraphics[scale=0.55]{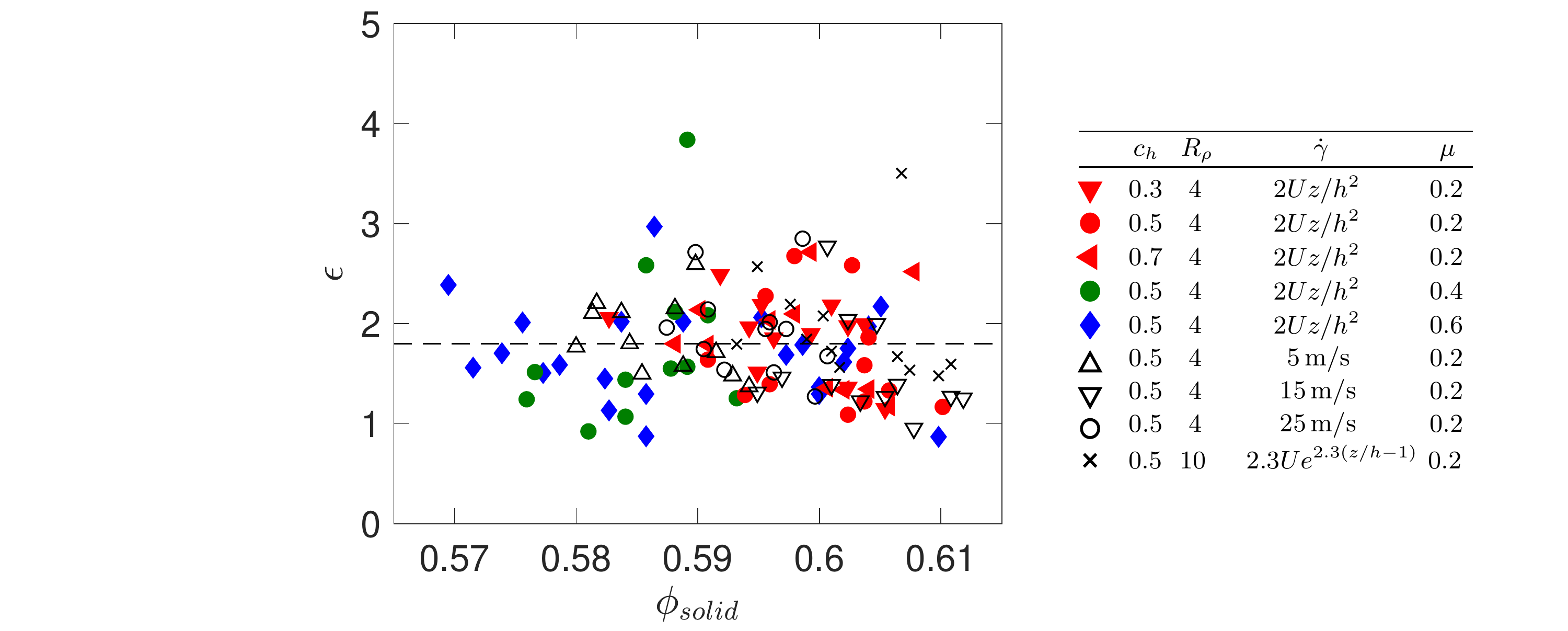}}
    \caption{Viscous drag coefficient $\epsilon$ calculated from Eq.~(\ref{stokesvell}) vs. solid volume fraction $\phi_{solid}$ for different flow configurations and friction coefficients. The dashed line indicates the mean value $\epsilon=1.73$.}
    \label{dragcoefficient}
\end{figure}

\begin{figure}
    \captionsetup{justification=raggedright}
    \centerline{\includegraphics[scale=0.55]{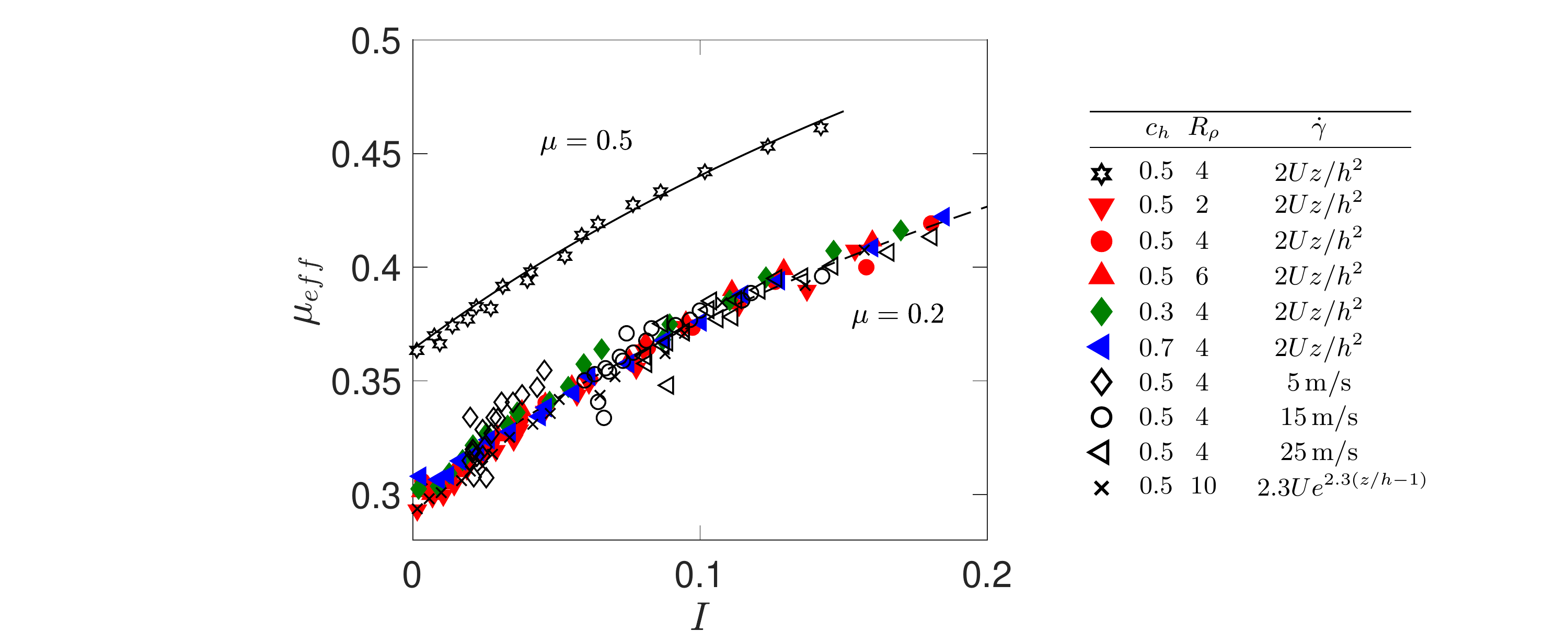}}
    \caption{Effective friction coefficient $\mu_{eff}$ vs. local inertial number $I$ for different combinations of density ratio $R_\rho$, heavy particle concentration $c_h$, and shear rate $\dot\gamma$ with $e=0.9$ and $U=2\,$m/s. Different colors and symbols represent simulations with $\mu=0.5$ (top) and $\mu=0.2$ (bottom). The solid curve shows the prediction of Eq.~(\ref{epsilon}) with $\mu_s=0.364$, $\mu_2=0.772$, and $I_c=0.434$ for data from a previous study on chute flow \cite{tripathi2011rheology}. The dashed curve shows the prediction of Eq.~(\ref{epsilon}) with $\mu_s=0.3$, $\mu_2=0.68$, and $I_c=0.4$ for data from this study. }
    \label{rheology}
\end{figure}

The viscous drag model of Eqs.~(\ref{stokesvell}) and (\ref{stokesvelh}) has previously been described in detail \cite{tripathi2013density}, so here we merely determine the coefficients $\epsilon$ and $\eta$ so we can apply the model for comparison to the KTGF segregation model. The shear stress $\tau$ and shear rate $\dot\gamma$ from the DEM simulations can be used to estimate the granular pseudo-viscosity $\eta$ ($\eta=\tau/\dot\gamma$). This in turn allows the calculation of the empirical viscous drag coefficient $\epsilon$ from Eqs.~(\ref{stokesvell}) and (\ref{stokesvelh}). 
Figure~\ref{dragcoefficient} plots $\epsilon$ versus solid volume fraction $\phi_{solid}$ for different flow configurations and friction coefficients, $\mu$. 
$\epsilon$ is relatively constant, having a value of about 1.7 for $0.57\leq \phi_{solid}\leq 0.61$.
This value is somewhat lower than that from a previous study \cite{tripathi2013density}, in which $\epsilon$ decreases from 3.7 to 2.5 as $\phi_{solid}$ increases from 0.51 to 0.58 for intruder particles in chute flow.

Equations~(\ref{stokesvell}) and (\ref{stokesvelh}) also require an expression for $\eta$, which can be written in terms of the effective friction coefficient $\mu_{eff}=\tau/P$, such that
\begin{equation}
	{\eta=\mu_{eff} \frac{P}{\dot\gamma}.
	\label{viscosity}
}\end{equation}
According to the $\mu(I)$ rheology \cite{jop2005crucial}, $\mu_{eff}$ is a function of the inertial number $I$;
\begin{equation}
	\mu_{eff}=\mu_{s}+\frac{\mu_2-\mu_s}{I_c/I+1}.
	\label{epsilon}
\end{equation}
The three rheological parameters $\mu_s$, $\mu_2$, and $I_c$ are material specific, and need to be independently determined for each value of interparticle $\mu$. 
The effective friction coefficient $\mu_{eff}$ estimated from DEM simulations ($\mu_{eff}=\tau/P$) is plotted as a function of $I$ for $\mu=0.2$ and 0.5 in Fig.~\ref{rheology}.  
For the set of data with $\mu=0.5$, we validate our calculation by comparing the simulation results to the predictions of Eq.~(\ref{epsilon}) with $\mu_s=0.364$, $\mu_2=0.772$, and $I_c=0.434$ from a previous study on chute flow \cite{tripathi2011rheology}. 
For the other set of data with $\mu=0.2$, Fig.~\ref{rheology} provides the relation between $\mu_{eff}$ and $I$ for different combinations of density ratio $R_\rho$, concentration $c_h$, shear rate $\dot\gamma$, and vertical location in the simulation.
The data collapse on a curve with some outliers for uniform shear flows due to the discontinuity in local shear rate near the wall.
$\mu_{eff}$ for $\mu=0.2$ is somewhat smaller than that for $\mu=0.5$, demonstrating the $\mu(I)$ rheology dependence on interparticle friction $\mu$.
This is expected, noting that $\mu_s$ is related to the inclination angle for steady chute flows. Increasing $\mu$ results in a higher inclination angle and hence larger $\mu_s$.
For $\mu=0.2$, Eq.~(\ref{epsilon}) with $\mu_s=0.3$, $\mu_2=0.68$, and $I_c=0.4$ agrees well with the data in Fig.~\ref{rheology}. We also observe that $R_\rho$, $c_h$, and $\dot\gamma$ have little influence on the $\mu(I)$ relation, consistent with previous results for chute flows \cite{tripathi2011rheology}.

\section{Segregation velocity model predictions}
\label{section5}

\begin{figure}
     \captionsetup[subfigure]{labelformat=empty}
    \captionsetup{justification=raggedright}
    \begin{subfigure}{0.37\columnwidth}
    \includegraphics[scale=0.5]{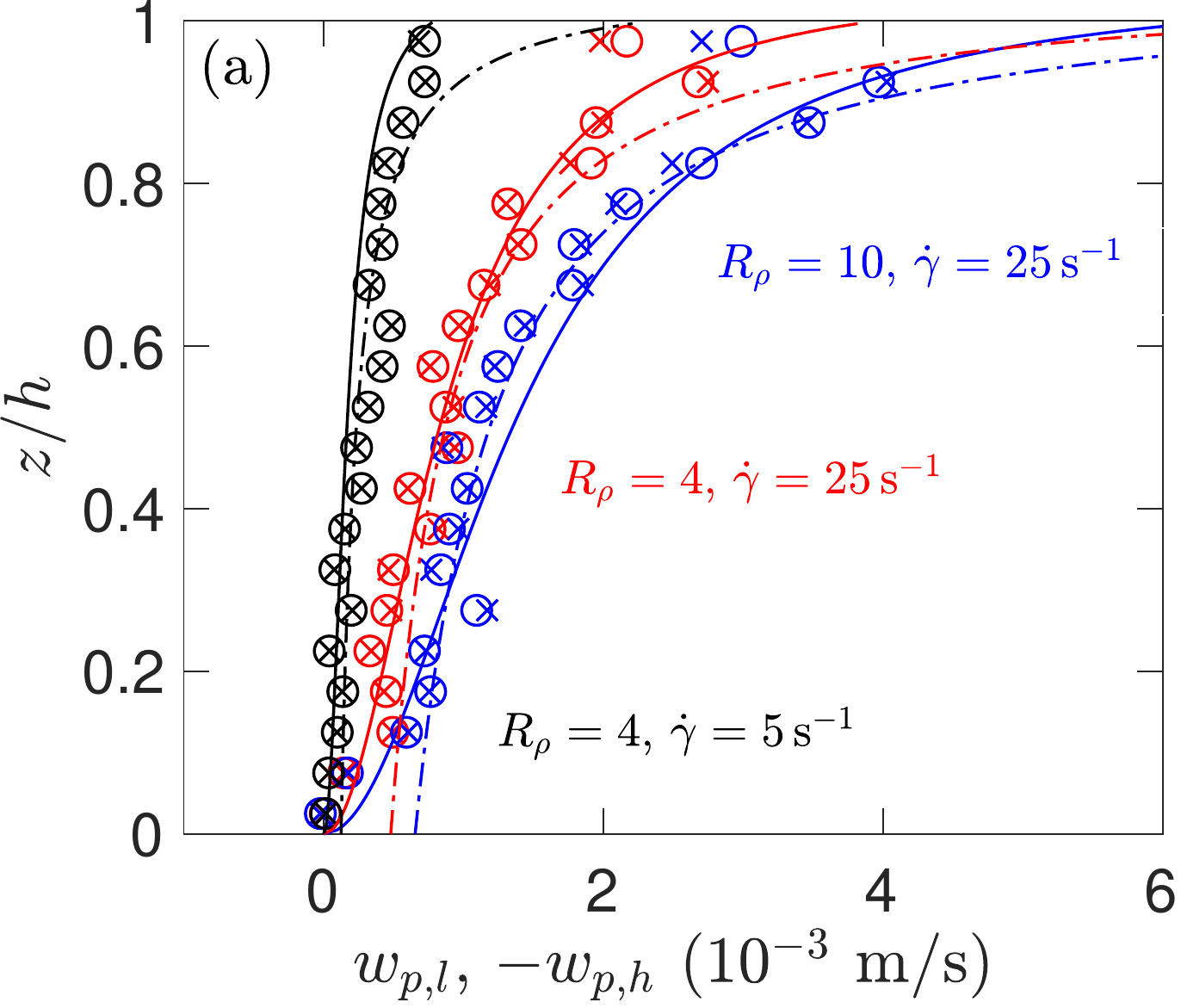}
    \subcaption{}\label{perco_profile1}
     \end{subfigure}
    \begin{subfigure}{0.43\columnwidth}
    \includegraphics[scale=0.5]{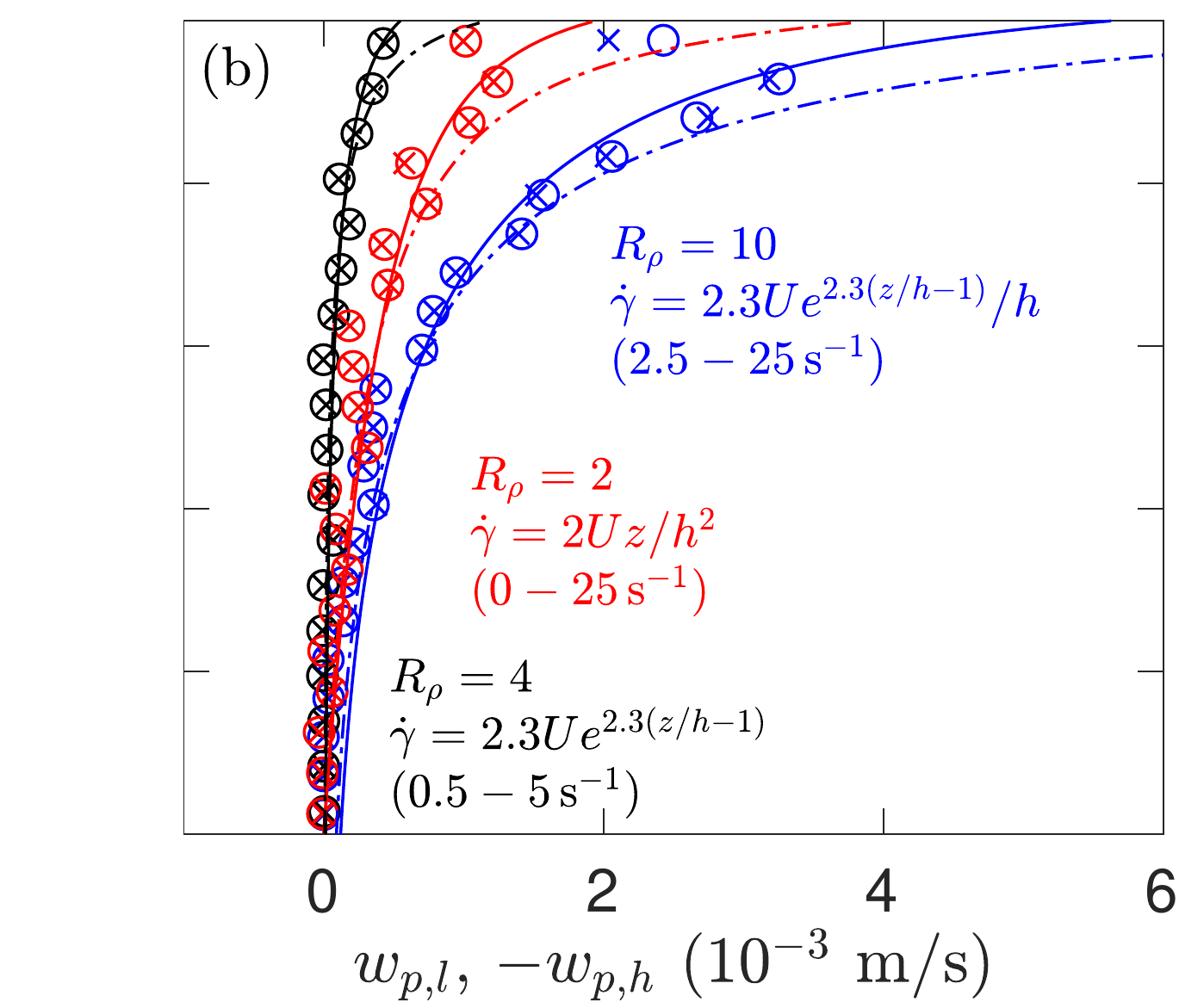}
    \subcaption{}\label{perco_profile2}
    \end{subfigure}    
    \caption{Predicted segregation velocity profiles from modified KTGF drag model [Eqs.~(\ref{final}) and (\ref{finalh})] (solid curves) and viscous drag model [Eqs.~(\ref{stokesvell}) and (\ref{stokesvelh})]  (dashed curves) compared to DEM simulation data for light ($\times$) and heavy ($\Circle$) particles under (a) uniform ($\dot\gamma=U/h$) and (b) varying shear rate profiles [$\dot\gamma=$$2Uz/h^2$, $2.3Ue^{2.3(z/h-1)}/h$] with different density ratios. $\mu=0.2$, $c_h=c_l=0.5$, and $e=0.9$. }
    \label{perco_profile}
\end{figure}

\begin{figure}
    \captionsetup{justification=raggedright}
    \begin{subfigure}{0.39\columnwidth}
    \includegraphics[scale=0.5]{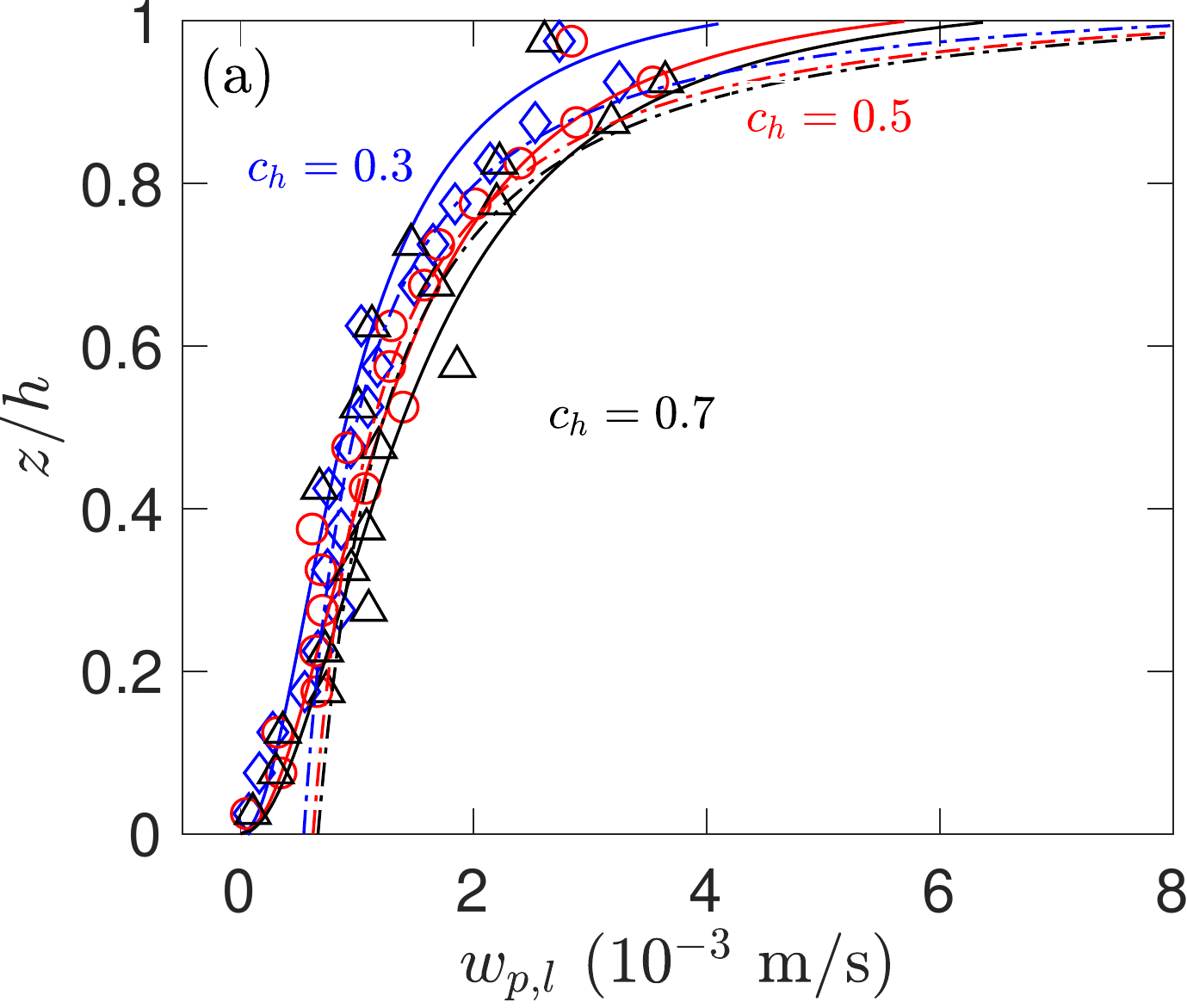}
    \label{wp}
     \end{subfigure}
    \begin{subfigure}{0.39\columnwidth}
    \includegraphics[scale=0.5]{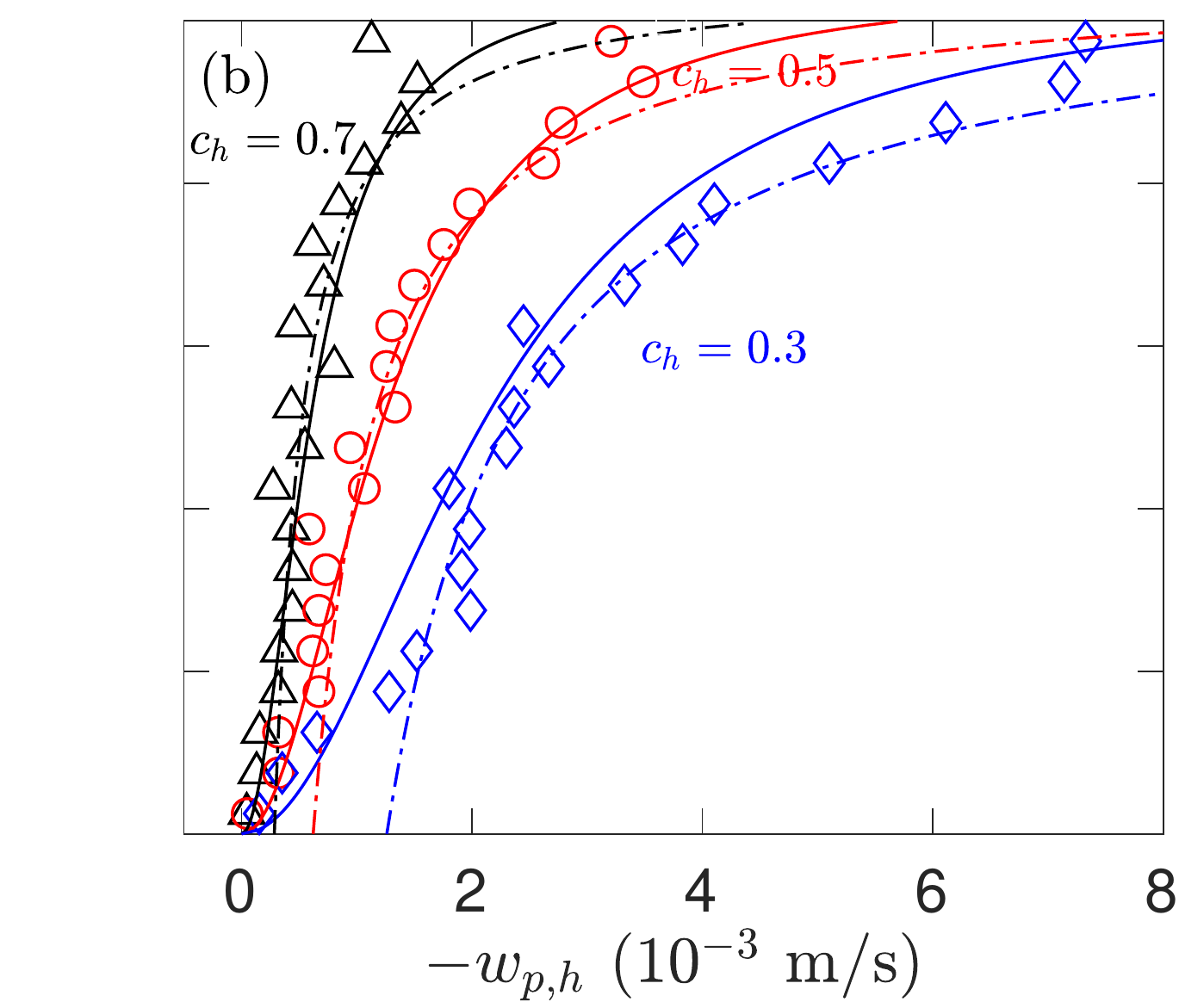}
    \label{i}
    \end{subfigure}
    \caption{ 
   Predicted segregation velocity profiles from modified KTGF drag model [Eqs.~(\ref{final}) and (\ref{finalh})] (solid curves) and viscous drag model [Eqs.~(\ref{stokesvell}) and (\ref{stokesvelh})]  (dashed curves) compared to DEM simulation data under uniform shear for (a) light and (b) heavy particles with heavy particle concentration $c_h=$0.3, 0.5 and 0.7 ($c_l=1-c_h$). $\mu=0.2$, $R_\rho=8$, $\dot\gamma=25~\text{s}^{-1}$, and $e=0.9$.}
    \label{shear}
\end{figure}

With the empirical terms of both segregation velocity models determined, we now compare the predictions of Eqs.~(\ref{final}), (\ref{finalh}) and Eqs.~(\ref{stokesvell}), (\ref{stokesvelh}) to the segregation velocities measured in the DEM simulations. For the KTGF drag model [Eqs.~(\ref{final}) and (\ref{finalh})], the global values for $d$, $\phi_{solid}$, $c_l$, $c_h$, $R_\rho$, and the empirically determined form of $B(\mu)$ are used in the model along with local values of $I$ at each depth in the flow. The local inertial number $I$ is based on the overburden pressure, estimated as $P=P_{wall}+\rho_{solid}\phi_{solid}g(h-z)$, such that $I$ can be expressed as a function of vertical position $z$ according to Eq.~(\ref{inert}) (or, equivalently, Eq.~(\ref{istar}) for uniform shear flows).
For the segregation model based on the modified Stokes viscous drag force [Eqs.~(\ref{stokesvell}) and (\ref{stokesvelh})], the global values for $d$, $c_l$, $c_h$, $\rho_l$, $\rho_h$, and the empirical value for $\epsilon$ are used along with the local values for $\eta$ based on the $\mu(I)$ rheology in Eq.~(\ref{epsilon}) at each depth.

Figure \ref{perco_profile1} compares the particle segregation velocity profiles predicted by the KTGF drag model [Eqs.~(\ref{final}) and (\ref{finalh})] and the viscous drag model [Eqs.~(\ref{stokesvell}) and (\ref{stokesvelh})] with those measured from DEM simulations of uniform shear flows in Fig.~\ref{wp-i}.
Both models represent the simulation data reasonably well.
Considering that the segregation velocities for $z/h<0.2$ and $z/h>0.8$ in the simulations are affected by the accumulation of segregating particles near the bounding walls.
The KTGF drag model seems to handle wall influence slightly better than the viscous drag model, most likely because of the use of the modified inertial number that accounts for the nonlocal effect of the boundaries. Without further modification to the $\mu(I)$ rheology, the viscous drag model does not include the nonlocal effect of the boundaries.

A more rigorous test of the model is to consider flows with varying shear profiles $[\dot\gamma=2Uz/h^2$, $2.3Ue^{2.3(z/h-1)}/h]$, see Fig.~\ref{perco_profile2}. 
The close match between the KTGF segregation model prediction and the data demonstrates that Eqs.~(\ref{final}) and (\ref{finalh}) are also effective for flow with non-uniform shear rates, even though $B(\mu)$ in the model was  derived from flow with uniform shear rate ($\dot\gamma=U/h$). 
The segregation model based on the viscous drag is also successful except near the upper wall due again to nonlocal effects.

Figure \ref{shear} provides a similar comparison where the concentration ratio ${c_h}/{c_l}$ is varied while keeping other parameters constant for uniform shear. 
In this case, the segregation velocities for heavy and light particles differ from one another when the concentration of the two species is unequal, as expected \cite{jones2018asymmetric}. 
Again, the estimated segregation velocities using both drag models match the DEM data reasonably well. The segregation model based on the viscous drag represents the near collapse of the data for $w_{p,l}$ in Fig.~\ref{shear}(a) slightly better than the KTGF segregation model. However, the KTGF segregation model works somewhat better for $w_{p,h}$ in Fig.~\ref{shear}(b), particularly near the upper and lower walls. Thus, we conclude that the new segregation model [Eqs.~(\ref{final}) and (\ref{finalh})] based on the interspecies drag term derived from analogy with KTGF can be used as an alternative to the viscous drag model to estimate the segregation velocities of both light and heavy species through the depth of the flowing layer under a variety of flow conditions.

  \begin{figure}
     \captionsetup[subfigure]{labelformat=empty}
    \captionsetup{justification=raggedright}
    \centering
    \begin{subfigure}{0.4\columnwidth}
    \includegraphics[scale=0.5]{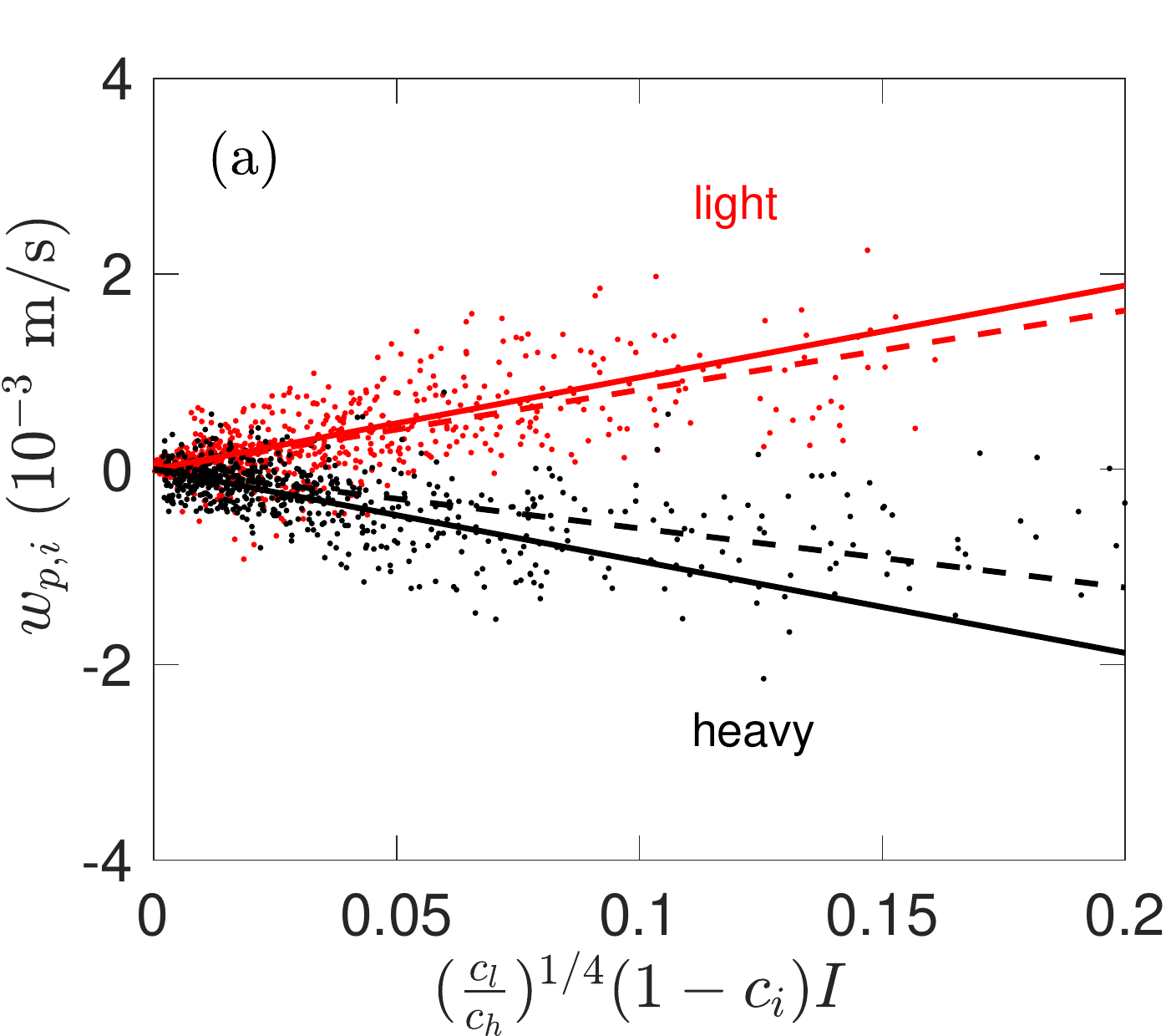}   
    \subcaption{}\label{heap1}
    \end{subfigure}
    \begin{subfigure}{0.39\columnwidth}
    \includegraphics[scale=0.5]{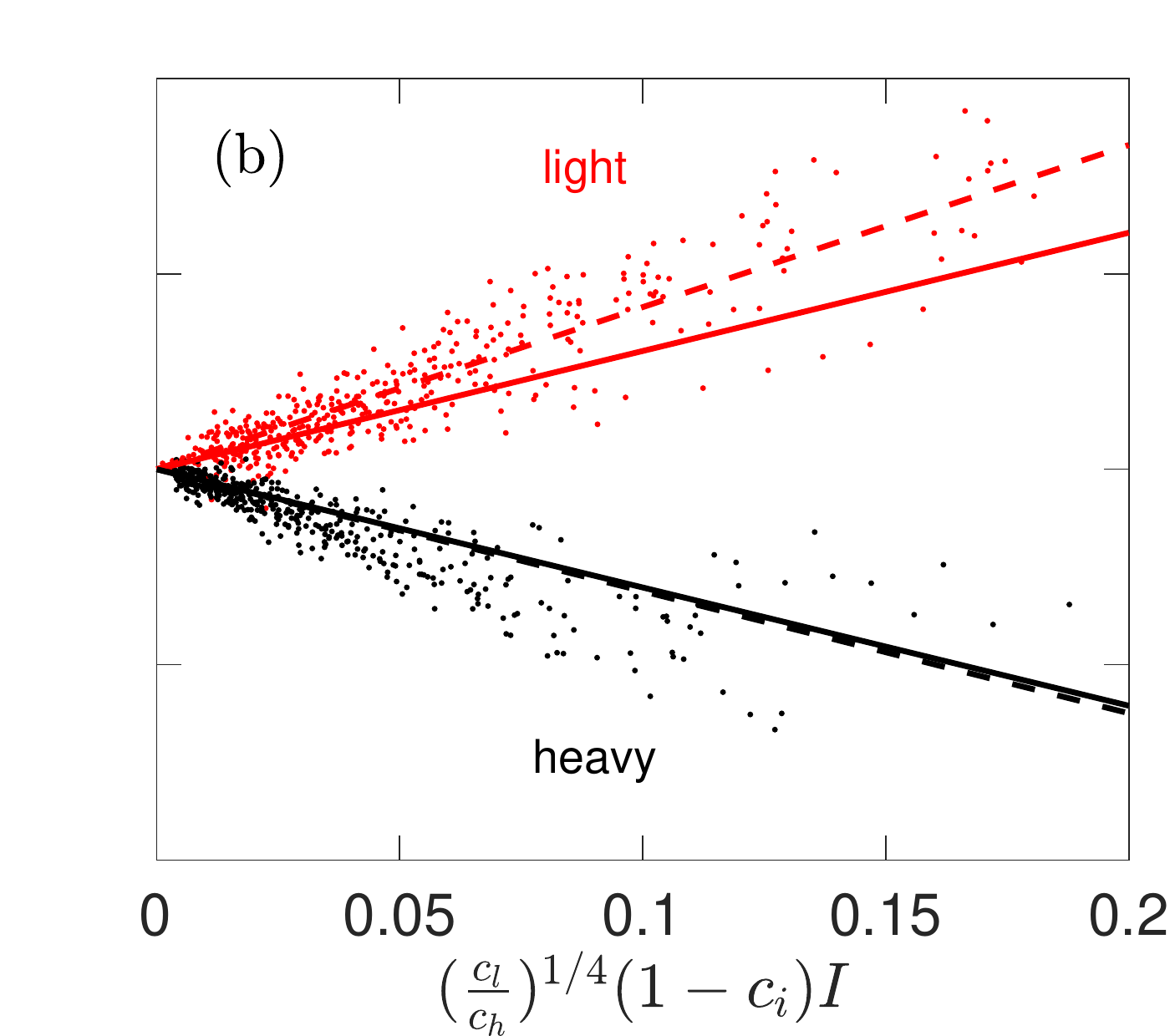}  
    \subcaption{}\label{heap2}
    \end{subfigure}
    \caption{Segregation velocities from DEM heap flow simulations (points) \cite{xiao2016modelling} for 
    (a) $R_\rho=1.84$, $\mu=0.2$, $e=0.9$, and $d=3~$mm, and (b) $R_\rho=3.33$, $\mu=0.2$, $e=0.9$, and $d=2~$mm.
    Dashed lines are linear fits to the data, and solid lines are predictions of the model based on the KTGF drag model. }
   \label{heap}
\end{figure}

Unlike the confined shear flows discussed to this point, the shear rate and the concentration ratio of many free surface flows vary across the domain. To further test the KTGF drag model under these more general conditions, we consider quasi-2d bounded one-sided heap flow in which particles flow in a thin surface layer down a slope much like what occurs when filling a silo \cite{fan2012stratification,fan2014modelling,deng2018continuum,deng2019modeling}.
Figure \ref{heap} shows segregation velocities from heap flow DEM simulations  \cite{xiao2016modelling} versus $({c_l}/{c_h})^{1/4}(1-c_i)I$, since, according to Eqs.~(\ref{final}) and (\ref{finalh}), these two variables are linearly related when all other conditions are equal.
The difference between these results and those presented up to this point is that these results correspond to a wide range of shear rates and concentrations, all occurring simultaneously in the flowing layer of the heap. 
The solid lines in Fig.~\ref{heap} are based on the KTGF drag model, the data points represent DEM results \cite{xiao2016modelling}, and the dashed lines represent least squares fits through the data points.
Overall, the segregation velocities of both bounded heap flows are reasonably well predicted by the model. 
However, since segregation in heap flows mainly occurs in a thin layer at the free surface, the averages are more uncertain, so the data have substantial scatter, especially for the weakest segregation case with $R_\rho=1.84$ and $d=3~$mm in Fig.~\ref{heap1}. The case with a small particle diameter $d=2~$mm but large density ratio $R_\rho=3.3$ has overall larger $w_{p,i}$ such that the DEM results match the theoretical predictions of Eqs.~(\ref{final}) and (\ref{finalh}) better, as shown in Fig.~\ref{heap2}. Note that we only include the heap flow data for $0.1 \le c_h,~ c_l \le 0.9$ to match the concentration range for flows in which our proposed KTGF drag model is valid. 

\begin{figure}
     \captionsetup[subfigure]{labelformat=empty}
    \captionsetup{justification=raggedright}
    \centering
    \begin{subfigure}{0.45\columnwidth}
    \includegraphics[scale=0.5]{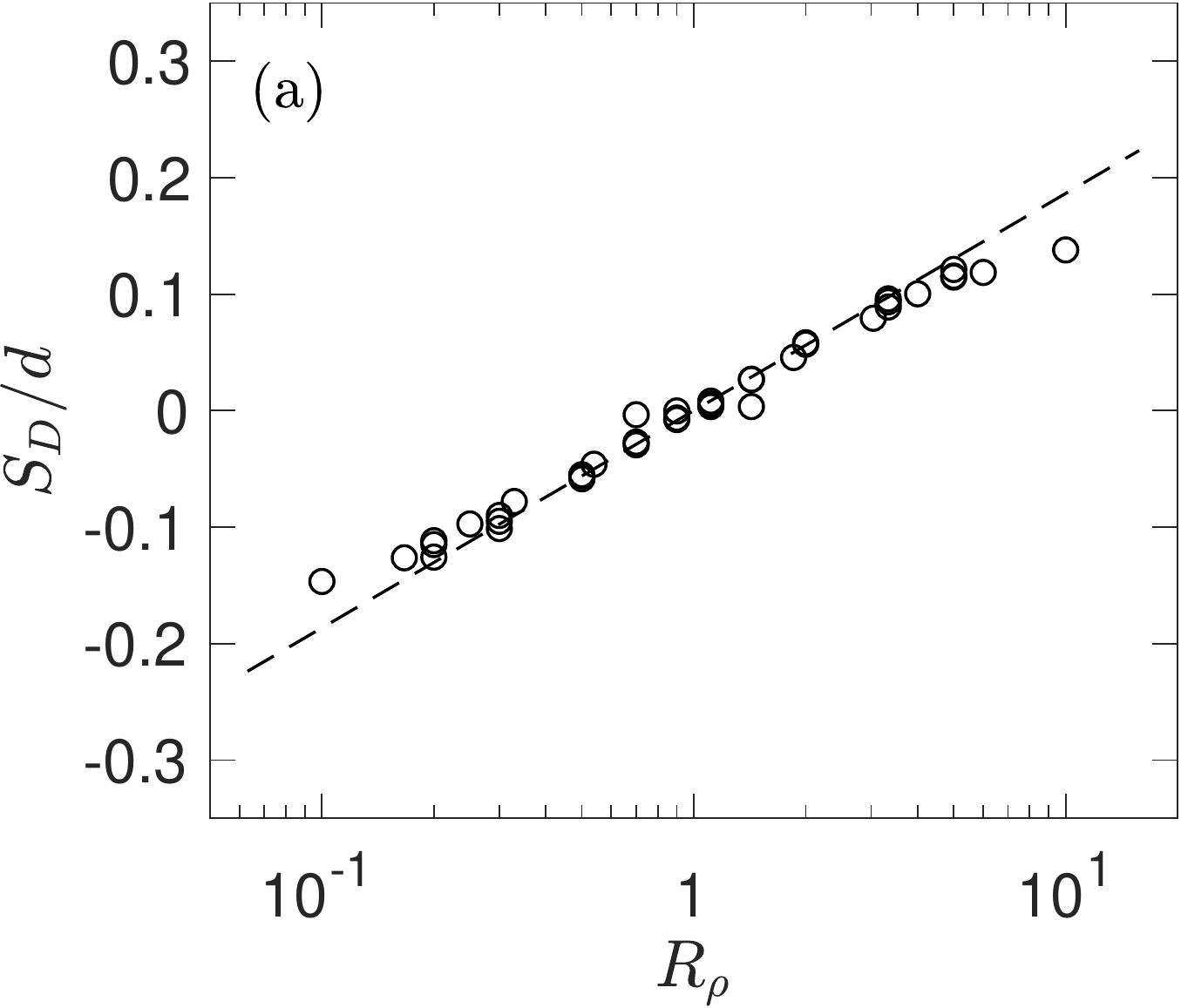}   
    \subcaption{}\label{hongyi1}
    \end{subfigure}
    \begin{subfigure}{0.45\columnwidth}
    \includegraphics[scale=0.5]{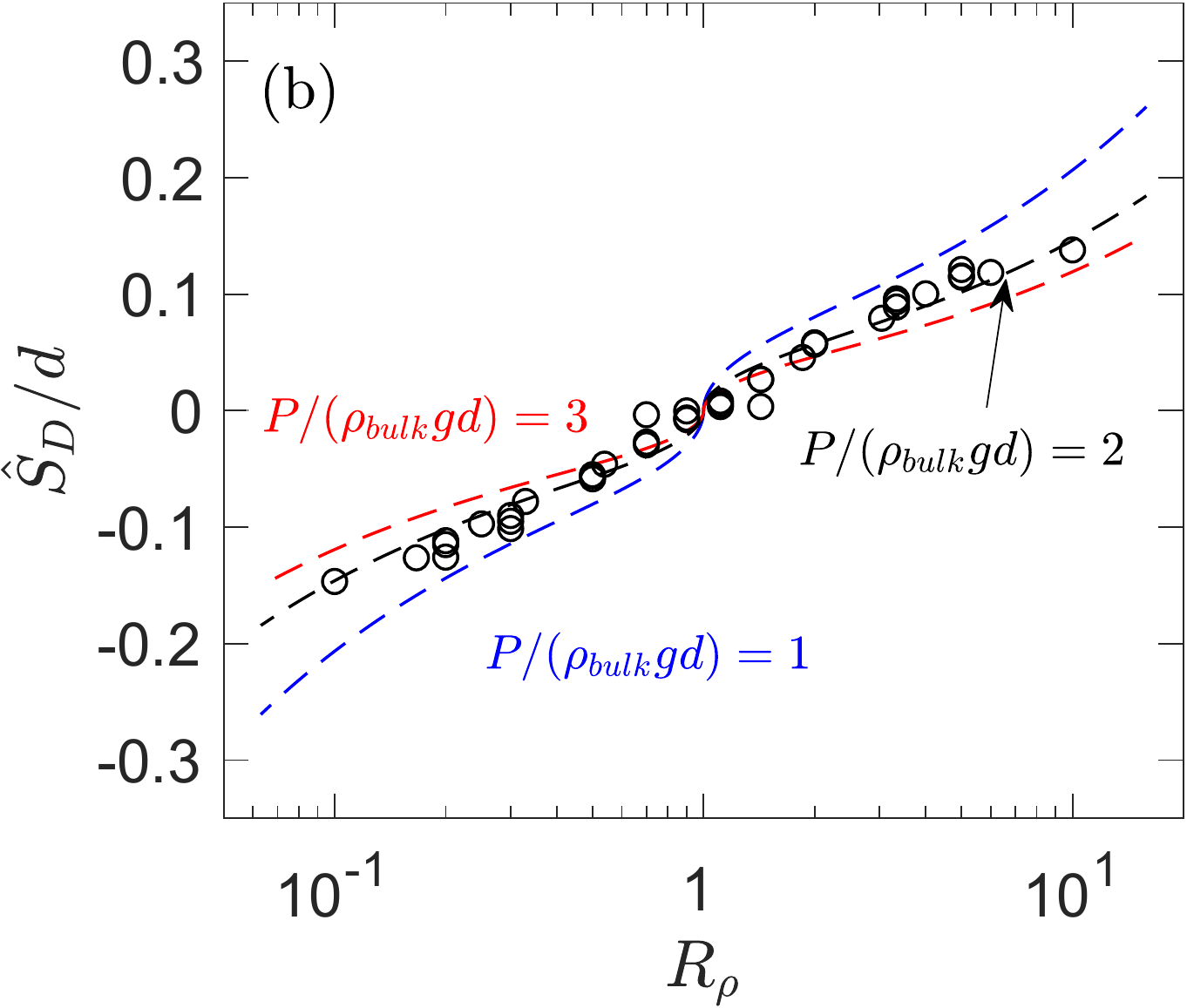}  
    \subcaption{}\label{hongyi2}
    \end{subfigure}
    \caption{Dependence of segregation length scale $S_D$ on the density ratio $R_\rho$ from DEM simulations (circles) for segregation in a bounded heap flow \cite{xiao2016modelling} compared to predictions of (a) the empirical fit of Eq.~(\ref{sd2}) and (b) Eq.~(\ref{sd1}) derived from the KTGF drag model with $c_h/c_l=1$, $\mu=0.2$ and different values of the mean pressure in the flowing layer $P/(\rho_{bulk}gd)$. }
   \label{hongyi}
\end{figure}

As Fig.~\ref{heap} shows, the segregation model based on the KTGF drag we propose in this paper works reasonably well not only for the confined shear flow between two planes from which it is developed, but also for free surface flow in a bounded heap.
Previous results \cite{xiao2016modelling} suggest that the segregation velocity can be modeled as
\begin{equation}
    |w_{p,i}|=\hat S_D  \dot\gamma(1-c_i) 
    \label{simplified}
\end{equation}
for bounded heap flow with inlet concentration $c_l=c_h=0.5$. Equations~(\ref{final}) and (\ref{finalh}) have this form when
$\hat S_D$, the segregation length scale, is
\begin{equation}
\frac{\hat S_D}{d}=\bigg[ \frac{1}{B(\mu) \phi_{solid}^2}   \Big( R_\rho-\frac{1}{R_\rho} \Big)  \sqrt{\frac{c_l}{c_h} } \Big( \frac{\rho_{bulk}gd}{P} \Big)   \bigg]^{1/2},
\label{sd1}
\end{equation}
where $\rho_{bulk}=\rho_{solid}\phi_{solid}$ is the bulk density.
For free surface flows, the empirically determined segregation length scale $S_D$ is well approximated by \cite{xiao2016modelling}
\begin{equation}
\frac{S_D}{d}=C_D \text{ln} R_\rho,
\label{sd2}
\end{equation}
where $C_D=0.081$. The empirical relation and the underlying data are shown in Fig.~\ref{hongyi1}.

To compare $S_D$ and $\hat S_D$ it is necessary to account for the pressure and particle concentration ratio, as well as other flow conditions. The comparison is made by first assuming ${{c_l}/{c_h} }=1$, which is the feed concentration ratio in these simulations, such that $\hat S_D$ is no longer concentration dependent.
Furthermore, in heap flows, segregation occurs in a thin flowing layer, with a thickness of only a few particle diameters, in which the local shear rate decreases exponentially with depth while the scaled local pressure $P/(\rho_{bulk}gd)$ increases linearly from 1 at the surface to $\delta/d$ at the bottom of the flowing layer at depth $\delta$, assuming constant $\rho_{bulk}$. 
The mean shear rate of the flowing layer equals the local shear rate at a depth of $0.26\delta$ due to the exponential streamwise velocity profile. 
This depth corresponds to approximately $2d$ for $7d\le \delta \le 8.5d$ \cite{xiao2016modelling}.
By assuming the exponentially varying local shear rate is the dominant factor for the depth varying segregation velocity, we expect that the effective pressure is equivalent to the pressure at a depth of about $2d$ such that $P/(\rho_{bulk}gd)=2$. 
For comparison, $P/(\rho_{bulk}gd)=1$ and 3 are also considered.

The expression for $\hat S_D$ from Eq.~(\ref{sd1}) for $c_h/c_l=1$ and three different pressures is compared to bounded heap flow results \cite{xiao2016modelling} in Fig.~\ref{hongyi2}. The data match the model curve well for $P/(\rho_{bulk}gh)=2$. 
Again, the results in Fig.~\ref{hongyi2} demonstrate the success of the KTGF drag model in replicating a wide range of results, even down to the slight curvature evident in the data matching that of the curve corresponding to Eq.~(\ref{sd2}).

As for the viscous drag model, it is possible to combine Eqs.~(\ref{stokesvell}), (\ref{stokesvelh}), (\ref{viscosity}), and (\ref{epsilon}) and rewrite the  expression for the segregation velocity,
\begin{equation}
	w_{visc,i}=\frac{gd^2}{6\epsilon}\frac{\rho_j-\rho_i}{P}\frac{1}{ \mu_s+\frac{\mu_2-\mu_s}{I_c/I+1} }(1-c_i)\dot\gamma.
	\label{wpexpand}
\end{equation}
Noting that Eq.~(\ref{wpexpand}) has the same form as Eq.~(\ref{simplified}) \cite{xiao2016modelling}, we can extract an expression for $\hat{S}_{D,visc}$,
\begin{equation}
	\frac{\hat S_{D,visc}}{d}=\frac{gd}{6\epsilon} \frac{|\rho_i-\rho_j|}{P}\frac{1}{ \mu_s+\frac{\mu_2-\mu_s}{I_c/I+1} }.
	\label{bothpandi}
\end{equation}
However, the resulting expression indicates that $\hat S_{D,visc}$ depends on pressure directly (inversely) and also through dependence on $I$, which depends on pressure. Assuming a simple linear dependence of pressure on depth ($P=\rho_{bulk} g z$) results in values of $\hat S_{D,visc}$ that are quite a bit larger than those that have been measured in DEM simulations of heap flow \cite{xiao2016modelling}. The reasons behind this are beyond the scope of this study, but may be related to applying parameters for the $\mu(I)$ rheology extracted from the shear flow in Fig.~\ref{scheme} to the free surface flow occurring on a bounded heap. In any case, it is difficulty to further compare $\hat S_{D,visc}$ to previous heap simulation data, as was done for the KTGF drag model in Fig.~\ref{hongyi}.

\section{conclusion}
\label{section6}

In this paper, we propose a semi-empirical predictive model, Eqs.~(\ref{final}) and (\ref{finalh}), for the segregation velocities of light and heavy particle species in density bidisperse granular flows that is based on a new model for the interspecies drag in segregating dense flows, Eq.~(\ref{eq8}).
The interspecies drag model assumes that the multiple long-duration particle interactions in dense granular flows reflect similar physics to short-duration binary particle interactions typical of dilute granular flows, which have been successfully modeled using the Kinetic Theory of Granular Flows (KTGF) \cite{jenkins1983theory,lun1984kinetic}. 
Of course, particle segregation depends on the pressure-shear state, which can be characterized by the inertial number $I$.
In particular, $I$ is inversely proportional to the square root of pressure, which can significantly reduce the segregation velocity \cite{fry2018effect}. 
In addition, the segregation depends non-linearly on the local particle concentration: heavy particles among many light particles segregate more quickly than light particles among many heavy particles \cite{jones2018asymmetric}.
Finally, particles in dense granular flows experience enduring contacts characterized by interparticle friction, which is in contrast to dilute granular flows where particle contacts are short and dominated by the elastic properties of the particles. Hence, the restitution coefficient $e$ has little influence on segregation in the dense flows considered here.

One advantage of the proposed KTGF drag model for dense granular flows [Eq.~(\ref{eq8})] is that it links interspecies drag to segregation velocity and includes the effects of local flow conditions, $I$, local concentration ratio, $c_h/c_l$, and particle properties including density, size, and surface friction, $\mu$.
The segregation velocities derived from combining the interspecies drag model with the equilibrium momentum balance equation match the segregation velocities determined from DEM simulations in both confined shear and free surface heap flows as well as the modified viscous drag force model proposed by \textcite{tripathi2013density}, which is analogous to the viscous force acting on a particle settling in a fluid.
Either approach allows calculation of segregation velocities through the depth of the flowing layer for different density ratios and relative constituent concentrations. 

Despite these advances, the KTGF drag model proposed here is not without drawbacks. The segregation velocities predicted from Eqs.~(\ref{final}) and (\ref{finalh}) for the KTGF drag model as well as those predicted from Eqs.~(\ref{stokesvell}) and (\ref{stokesvelh}) for the viscous drag model, while showing the right trends, are only reasonable estimates, as is evident from the scatter in the DEM simulation data from the model predictions in Figs.~\ref{perco_profile}-\ref{heap}. 
This is likely a result of the many variables in the problem, as well as the stochastic nature of the forces on individual particles that drive the segregation and the concurrent collisional diffusion, which is not taken into account in either approach. 
Furthermore, both approaches require empirical inputs. The KTGF drag model requires an empirical relation, $B(\mu)$, that reflects dependence on interparticle friction; the viscous drag approach requires an empirical value for the coefficient of drag, $\epsilon$, as well as appropriate values for the $\mu(I)$ rheology, which, in turn, depend on the interparticle friction. Thus, in either case, the underlying challenge lies in relating interparticle friction to the rheology of the flow.

Nevertheless, the results in Fig.~\ref{hongyi} demonstrate a remarkable correspondence between the segregation velocity model and the simulation data for the KTGF drag model, particularly since the data are for a different flow than that from which the model was derived.
More research is needed under even more widely varying conditions to refine both models, particularly with respect to the influence of interparticle friction. Furthermore, both models are limited to density bidisperse granular materials. A more challenging problem is to connect the interspecies drag to the segregation velocity for size-bidisperse particles, or, even more difficult, particles that differ in both size and density.

\nocite{*}

\bibliography{apssamp}
\bibliographystyle{apsrev4-2}
\end{document}